\documentclass{article}
\usepackage[utf8]{inputenc}
\usepackage{arxiv}
\usepackage{booktabs}
\usepackage{url}
\usepackage{caption}
\usepackage{subcaption}
\usepackage{graphicx} 
\usepackage{multicol}
\usepackage{amsmath}
\usepackage{rotating}
\usepackage{float}

\title{A Vulnerability Study on Academic Collaboration Networks Based on Network Dynamics}

\author{
    Asier Gutiérrez-Fandiño\\
    Barcelona Supercomputing Center\\
    \texttt{asier.gutierrez@bsc.es}\\
    \And
    Jordi Armengol-Estapé\\
    Barcelona Supercomputing Center\\
    \texttt{jordi.armengol@bsc.es}\\
    \And
    Marta Villegas\\
    Barcelona Supercomputing Center\\
    \texttt{marta.villegas@bsc.es}\\
}

\begin{document}
\maketitle

\begin{abstract}
On the one hand, researchers that work for the same institution use their email as the main communication tool. On the other hand, email can be one of the most fruitful attack vectors of research institutions as they also contain access to all accounts and thus to all private information.

We propose an approach for analyzing in terms of security research institutions' communication networks. Specifically, we suggest a method for reconstructing these networks as well as an approach to analyze possible breaches of collected emails, using the mentioned networks. For the former, we downloaded the network of 4 different research centers, three from Spain and one from Portugal. For the latter, we ran simulations of Susceptible-Exposed-Infected-Recovered (SEIR) complex network dynamics model for analyzing the vulnerability of the network.

In our simulations, we found that more than half of the studied nodes have more than one security breach. The results suggest that more than 90\% of the networks' nodes are vulnerable.

This method can be employed for enhancing security of research centers and can make email accounts users security-aware. It may additionally open new research lines in communication security. Finally, we manifest that, to prevent potential misuses, the sources we utilized for obtaining communication networks should reconsider providing the information that we were able to gather.

\end{abstract}


\section{Introduction}
In Information and Systems Security, maintaining the confidentiality of services' work is crucial, since it is the first defensive barrier to avoid unwanted access. Companies keep it confidential not only to avoid unwanted access, but also not to unveil how their processes and interactions work.

In the academic field, universities and research centers tend to be more open in revealing the way in which some processes work, personnel is organized and the involvement of processes and staff is incorporated, just to mention a few. This can be dangerous as universities and research centers are involved in relevant international research studies with, probably, sensitive data. Many cyberattacks attempting to steal the COVID-19 vaccine have occurred lately \cite{bbc_attack}\cite{nyt_attack}, which recalls that safeguarding researcher centers' security remains imperative.

Interestingly enough, it is a widely normalized practice to publish papers showing the email of the authors. In this paper, we show that using the same email address for public contact, services, and internal communications is a wrong practice as it can be treated as an attack vector used to smartly attack the whole institution. Using a web scraping technique to gather critical information of the institutions' researchers in the open web, we ensemble graphs of research interactions — namely \textit{Collaboration Networks} —, we search for the services employed by the researchers, and we perform simulations of attacks to these institutions.

\section{Related Work}

\cite{10.1145/586110.586140} use graphs for vulnerability analysis assessment. They build graphs with post and preconditions and elaborate an algorithmic model to reach these vulnerabilities. Similarly, \cite{jajodia2005} propose an approach close to the one from Ammann et al. They design an architecture based on obtaining data from novel vulnerabilities on the internet. Then, they store the knowledge about those vulnerabilities and, given a network, they determine the different attack paths. Our approach is  similar. We build a graph of the target network of communications and we look for vulnerabilities on the internet. Unlike Jajodia et al. and Ammann et al., though, we determine the potential attacks by simulations. Crucially, we provide an example with real data.

Many works analyzing academic networks have been proposed. \cite{10.3389/fphy.2016.00049} obtain citation graphs to find hidden relationships that they call 'cartels'. They express logically in a query format what they mean by a 'cartel' and they search for them in a database built by them. \cite{Zhao2019} present the APR framework for author-citation networks to measure paper influences and they test it against the Microsoft Academic Graph. Our paper proposes an innovative approach to obtain connections among researchers of a given institution and focuses on cybersecurity. 

Complex Network theory has been applied to cybersecurity for a long time. \cite{Shao_2015} analyze how a network evolves in terms of percolation in the context of an attack. \cite{Wen2017} introduce security recommendations for both cyber and physical networks. Both papers provide simulations to back their work.

\cite{10.1145/2600176.2600190} briefly introduce the idea of dynamics for cybersecurity, and provides as an example that nodes state changes over time. Other authors use a well-known family of network dynamics especially thought for diseases that fit cybersecurity requirements as well. \cite{10.1007/978-3-319-47364-2_63} perform a complete review of the different propagation models applied to cybersecurity. We refer to this work for the description of the concepts of Susceptible-Infected-Susceptible (SIS) and Susceptible-Exposed-Infected-Removed (SEIR) \cite{ARON1984665} as well. These models are Disease Propagation models based on Network Dynamics for graphs which work as conditional probabilistic state transitions.

\cite{8320357} use the SIS model to analyze campus network security, whereas, as in our paper, \cite{batista2018} employ the SEIR model to simulate computer virus spreading. \cite{zhu2019} follow the same model as \cite{8320357} for simulating virus spreading in smart grid networks.

Some authors, due to the special requirements of the security analysis they where analyzing, created models derived from the family of the SIS and SEIR models. \cite{ojha2019} propose the $SE_{1}E_{2}IRV$ model for simulating worm attack in wireless sensor networks, and \cite{AHN20154121} propose an C-SEIRA model for code infection in computer networks. \cite{LIU2016249} developed a new model for mobile malware propagation, which is slightly far from SIS and SEIR models. While these models are promising with regard to particularity and specificity, we do not find it necessary to introduce a new model; hence, we continue with the SEIR mode as it already meets our modeling assumptions and we do not need the aforementioned particularity and specificity.

\section{Methods}
We perform the analysis with the Python programming language and we open source our tools for research purposes \cite{code}. The code can be used to effectively check the vulnerability level of a research institution. It uses the NDLib library \cite{Rossetti_2017}, which implements the Network Dynamic simulation method, the one we employ.

\subsection{Obtaining the Collaboration Network}
\label{sec:col-net}
In order to construct the \textit{Collaboration Network}, we use the scraping technique over the Google Scholar website \cite{scholar}. We search for the email addresses of a certain institution and it responds with entries to papers with descriptions showing the email addresses of the researchers.

We define the \textit{Collaboration Network} as an undirected graph $G = (V,E)$ where vertices are researchers' emails and edges are the collaboration between two researchers weighted with the total number of collaborations performed in different papers. For simplicity and to focus only on the target institution, we limit the collaborations to the domain of the institution. In other words, researchers' emails with domains that are outside the institution are discarded. Scraping the relationships results in disconnected subgraphs $H_{i}$ s.t. $V(H_{i}) \subseteq V(G) \wedge E(H_{i}) \subseteq E(G)$ where $i$ is the index of a vector of subgraphs $[H_{0}, H_{1}, ..., H_{n}]$ sorted in descendent order from $|V(H_{i})|$. We take $H_{0}$ as our new graph $G$ for simplicity as we want to avoid disconnected subgraphs. Another approach could be linking the most popular nodes of each $H_{i}$ weighting with small $\lambda$; yet, we did not experiment with it.

\subsection{Obtaining Breaches}
We discover the breaches of the email addresses by using the Have I Been Pwned service \cite{haveibeenpwned}. This service lets users search for any email address and check which are the breaches that the address has previously had. We attach this information to each node as an attribute in the graph $G$, as a count of breaches.

Furthermore, with the aforementioned service we can know which services the researcher uses with the given email account. This is useful not only to test old passwords but also to obtain a significant amount of information about the target for an intelligent phishing attack.

\subsection{Characterizing the Network}
Prior to simulating the attacks, we get insights about the target network. We can characterize relevant nodes to which an attack would be more successful by using Kruskal's Minimum Spanning Tree (MST) algorithm \cite{kruskal1956shortest}, reversed to obtain the Maximum Spanning Tree. The MST algorithm finds a subset of edges of a connected undirected graph; hence, it acyclically links all vertices together and minimizing total edge weight. This algorithm will show the backbone structure of the network with, generally, the most relevant node in the center. Maximum Spanning Trees of the networks are attached in the Additional Material.

The Community Structure of the network shows clusters of communications, and plots of detections are included in the Additional Material. We use the Fluid Community Detection algorithm \cite{pares2017fluid} with $k = 10$ communities to detect. This information could be useful to attack clusters at the same time, or even to attack clusters separately with a different approach for each of them.

\subsection{Simulating the Attacks}
We use the Susceptible-Exposed-Infected-Recovered model as we have nodes that are susceptible and nodes that have already been exposed. Exposed nodes will eventually get hacked in the attack and then transform into infected nodes. Infected nodes that recovered the control of their hacked account would make a transition to Recovered. The SEIR model allows us to effectively represent these state transitions. The most basic property is that the nodes should be in either one of the four states; summing all proportions of nodes in each state results in 1. In the SEIR model, Exposed nodes are simulated as being already "infected", but they are immersed in an incubation period.

We define a threshold $T = 1$. For starting the simulation, nodes with a number of breaches higher than $T$ will start as Exposed, while the rest will start as Susceptible. We randomly select one Exposed node as Infected, and then, we sample for 40\% of the exposed nodes to be infected. The motivation behind this initial setting is that information about the nodes have been public for a long period of time and, thus, the start of the simulations should contain nodes already in the Infected state. Recall that exposed nodes are immersed in an incubation period so they will transition into infected eventually, thus, the percentage for transition sampling from Exposed to Infected has small impact. None of the nodes will start as Recovered.

We set the latent period of Exposed to Infected to be $\alpha = 0.05$, which implies a fast transition, and the infection probability to a large number $\beta = 0.2$, since hacked accounts could send e-mails to susceptible nodes and trust in colleagues would make nodes get hacked. Finally, the Infected to Removed transition probability is set to be even larger $\gamma = 0.3$, as the hacked accounts would get alerted by colleagues or noticed by themselves rapidly. We run the simulations 10 times for 80 iterations. The number of iterations that we define would represent the attack time window.

\section{Analysis}
For validating our method with real-life data, we selected universities with different characteristics to which the experiments apply, that is, public or private universities of a rather national or international standing and from different countries.

We decided to run the experiments with the University of Deusto \cite{deusto} (Deusto), an Spanish private university in Bilbao; University of Lisboa \cite{ulisboa} (ULisboa), a public university; Pompeu Fabra University \cite{upf} (UPF), a Spanish public university located in Barcelona; and Rovira i Virgili University \cite{URV} (URV), a Spanish public university located in Tarragona. Additionally, we ran the experiments with two research institutions involved in COVID-19 vaccines: the Nuffield Department of Medicine (NDM), from the Oxford University, and the Spanish National Biotechnology Center (CNB-CSIC). These experiments are attached in the Additional Material.

Table \ref{tab:university-metrics} shows the metrics of the graphs obtained by scraping. Universities with a larger number of nodes have a larger number of edges. The clustering coefficient of the smallest network in both the number of edges and the number of nodes is larger. The average shortest path in the smallest network is close to the one of the second smallest network and is larger to the one in the second largest network.

\begin{table*}[ht]
\centering
\begin{tabular}{@{}lrrrrr@{}}
\toprule
 & \multicolumn{1}{l}{N. nodes} & \multicolumn{1}{l}{N. edges} & \multicolumn{1}{l}{Avg. clustering coeff.} & \multicolumn{1}{l}{Avg. shortest path} & \multicolumn{1}{l}{Deg. assortativity coeff.} \\ \midrule
Deusto & 96 & 639 & 0.1881 & 2.8070 & 0.4631 \\
ULisboa & 214 & 3,505 & 0.0903 & 2.4679 & 0.4826 \\
UPF & 150 & 1,454 & 0.0450 & 2.9428 & 0.5377 \\
URV & 369 & 3,702 & 0.0077 & 4.2027 & 0.5944 \\ \bottomrule
\end{tabular}
\caption{Collaboration Network graph metrics.}
\label{tab:university-metrics}
\end{table*}

\begin{figure*}[!ht]
\centering
\includegraphics[width=12cm]{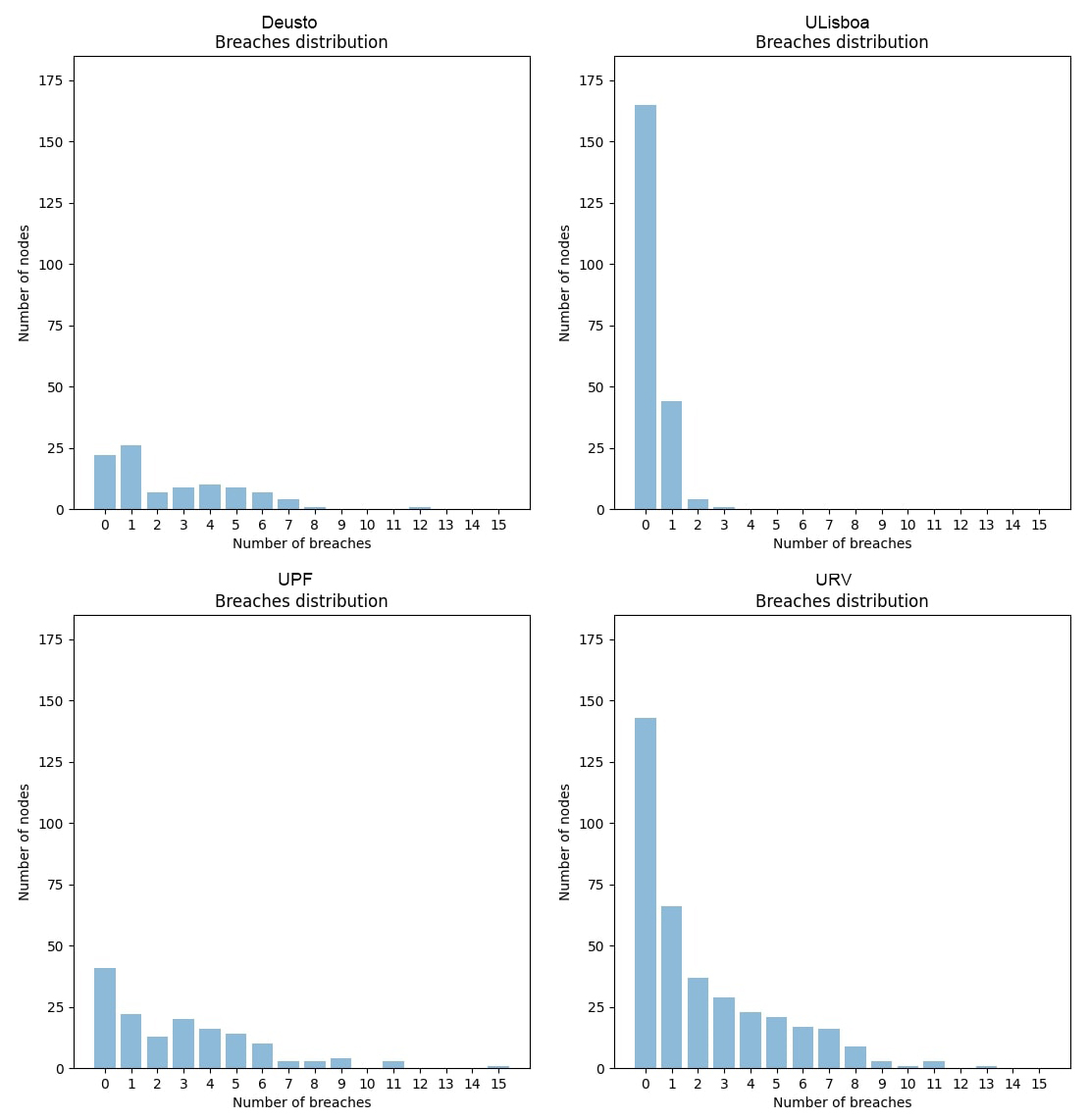}
\caption{Breach distribution bar plot.}
\label{fig:breach_barplot}
\end{figure*}

Breach distribution bar plots are shown in Figure \ref{fig:breach_barplot}. The distribution of the smallest network is flat, some of the nodes reaching 12 breaches; it is more homogeneously distributed than the rest. The second largest network from the figure located on top right only reaches 3 breaches; most of the nodes have 0 breaches. The second smallest network shows a homogeneously distributed breach distribution with a node reaching 15 breaches; the top number of breaches is 0 but it is close to the rest of the bars. The largest network shows clearly a descendant distribution from the peak of 0 breaches that is located in around 150 nodes; the tail of the distribution reaches 13 breaches. All these bar plots are worrying since, except in the case of the top right, most of the nodes have at least one breach or more. 

\section{Results}
\begin{figure*}[!ht]
\centering
\includegraphics[width=17cm]{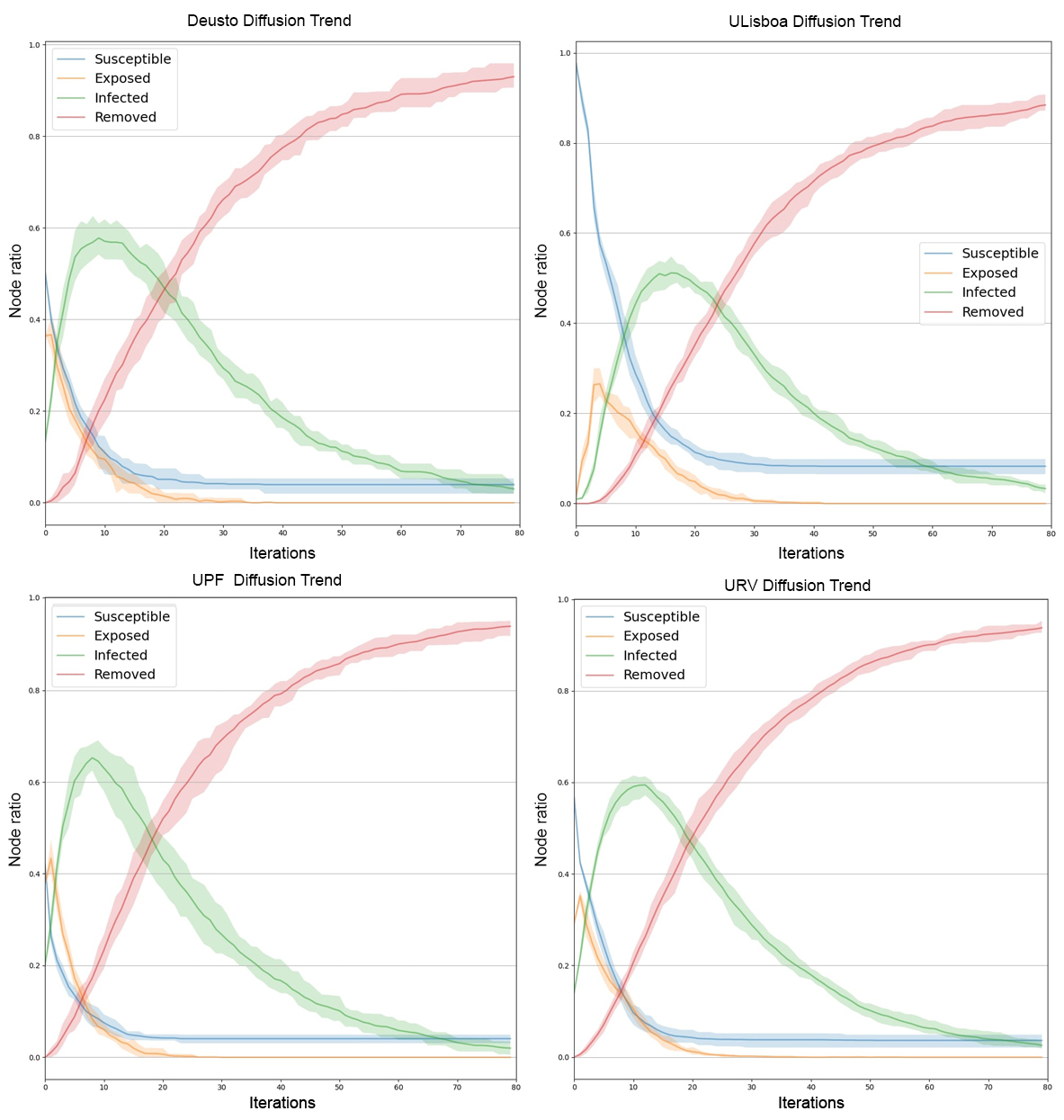}
\caption{Diffusion plot of SEIR model average of 10 runs, executed for 80 iterations.}
\label{fig:diffusion_trends}
\end{figure*}

Diffusion plots are shown in Figure \ref{fig:diffusion_trends}. The university with less breaches that is ULisboa, has the lowest exposure at the beginning, which is the one from the top right. ULisboa starts with 5\% of exposure, whereas the others start with 35-40\%. However, at the end of the simulations, ULisboa ends with around 10\% not being infected; this is the largest immunity percentage among the universities. The rest of the universities are close to the highest immunity level with the minimum in around 4\%. All percentages are low as they show that at least more than 90\% of the nodes get infected.

The UPF evidences a rather chaotic behaviour in the iterations 10-14 showing that more than the 60\% of the nodes will be infected at the same time. Close to this chaos is Deusto, indicating that some of the runs may reach the 60\%, as well as the university from the bottom right, which could touch this rate.

The red line corresponding to the Removed rate is similar in all the plots but smoother in the case of URV. The most singular case is the one from the ULisboa: it starts increasing after some iterations instead of increasing from the very beginning and it ends in a lower position than in the rest of plots.

\begin{figure*}[!ht]
\centering
\includegraphics[width=11cm]{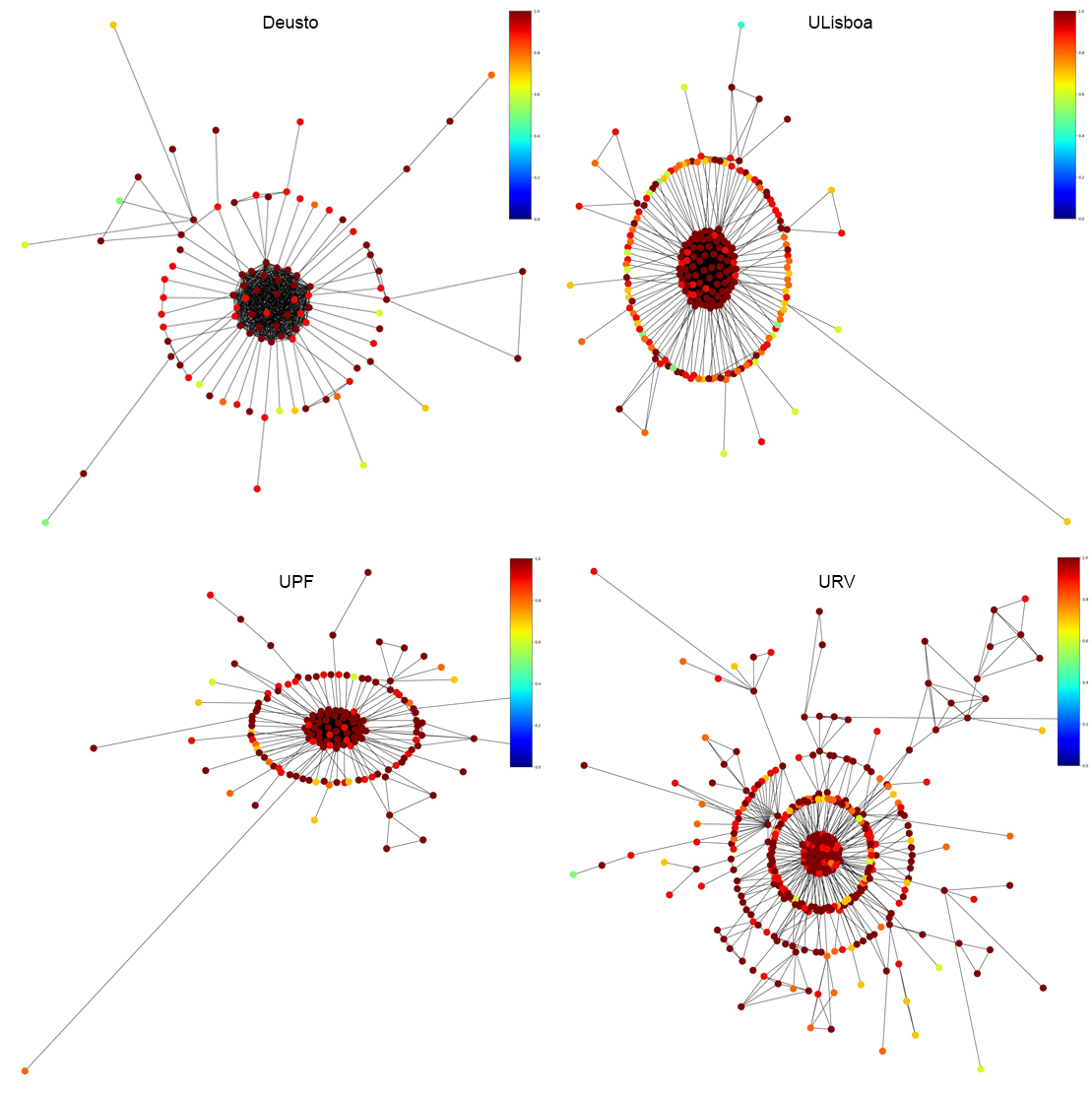}
\caption{Plots of the graphs using Kamada-Kawai algorithm \cite{kamada1989algorithm}. Colors are assigned according to node vulnerability level.}
\label{fig:graphs}
\end{figure*}

Figure \ref{fig:graphs} shows node infections of the networks; colors are assigned according to the colormaps presented in the plots. These colormaps are normalized so that the ratios of values are from 0 to 1. For instance, if a node has been infected in the ten simulations, it will have a ratio of 1 and, therefore, it will show garnet color; if a node has been infected half of the simulations, it will be 0.5 and it will show greenish color. Most of the nodes get infected at least once as the plot suggests. This plot is useful to understand the complexity of the graph. Deusto presents the simplest and flattest structure as it has only two levels, whilst the one from URV is a more complex one, showing up to 4 levels and some deep structures. In different structures the results are similar. Nodes from the center that are largely connected get infected more often than those that are on the outer structures.

\section{Discussion}
The information of the breaches may include previous passwords, name, surname and age, for instance. In addition, the information from the network reveals the behaviour of communication among colleagues. This could help attackers perform intelligent attacks with an increased success rate. Our simulations provide worrying results that corroborate our concerns.

The propagation of the attack that we model could be related to impersonating users and sending emails for phishing purposes. This phishing could have an even larger success rate as it would be sent internally (trust is larger). Furthermore, phishing attacks could be addressed from an account of a supervisor user, a research department or a unit providing supplementary success rate.

This simulation assumes that users get noticed from attacks and fix their vulnerabilities. Nevertheless, if an attack is made with sufficient discretion, it could result unnoticed by users and propagated through the network; consequently, spying could be conducted comfortably.

The impact on the research institution can be large since research could be hindered, be a victim of ransomware, or even stolen. Moreover, some research projects have private information for the studies; this information could be sensitive and, accordingly, should be kept secret. Information of this kind would also result affected.

We deeply discussed the publishing of this paper in terms of ethics. We encountered, among other reasons, that, if we have been able to discover this vulnerability pipeline, others could be able too. Making this vulnerability pipeline available as a supporting tool for security staff can be useful for determining whether personnel email is being used unintendedly; as we show in our paper, the more personnel email is misused, the worse the simulation results in terms of assessment are. Workers in charge of the security of the institution could obtain the vulnerability status in a realistic manner and apply the corresponding measures. Releasing the tool for research purposes may derive in enhancing our findings, creating new interesting works and improving the security of researchers.

Some of the target universities chosen for this study have not established the Two Factor Authentication (2FA) as mandatory. This fact, combined with the use of old passwords that are in databases of hackers, could be terrible since it makes trivial to steal an account; besides, getting the password stolen owing to a phishing attack would result in an account theft, too. Note that this network analysis is not only focused on this kind of attacks; these are mentioned because they are the most known ones due to their success rate.

We ask ourselves whether Google and other search services should avoid the indexing of email addresses, especially the ones from research institutions as it facilitates the scraping of \textit{Collaboration Networks}. This makes one of the keys in security — secrecy — vulnerated.

The scheme on how Have I Been Pwned website works might be even illegal as a whole, at least for European Citizens under the General Data Protection Regulation (GDPR). Trivially, by introducing an email address and without any identity verification nor even a captcha, all the breaches and services for the given person are unveiled. Most of the email addresses present in that database did not allow their emails to be there. We encourage these services to take further security measurements and, at least, require a verification code sent to the target email address for showing the corresponding information.

We strongly recommend researchers to avoid the use of the same communication address for internal and work communication, as well as for paper-related communication. We propose using alias accounts as a code (e.g. abc@institution.edu) or aliases linked to the papers; this way, communications can be straightforwardly identified by the paper (e.g. asier-vulnerability-study@institution.edu). Alternatively, 

\section{Conclusions and Future Work}
This study shows how communication networks of a given research institution can be approximated through collaborations in research papers. We additionally show how to enrich the network information to introduce possible breaches from other services linked to the email addresses. This enrichment could enhance the attacks as these breaches are from different databases; a large part of these databases has been published, while the other part is in the hands of bad actors.


Our results remain consistent in the simulations of different types of universities, either public or private; more than 90\% of nodes are highly vulnerable. The types of attacks that we consider according to the vulnerabilities presented and the simulation model selected are \textbf{phishing}, \textbf{account theft} and \textbf{spying}. Since the breaches provide private information about the user, directed attacks from an impostor service utilizing that information will logically have a larger success rate. Breaches containing passwords could be useful if users employ the same password across multiple services.


%

The procedure on which \textit{Collaboration Network} is obtained, which is explained in \ref{sec:col-net}, mentions that the method limits the email domains to the target institution. This restricts the propagation of an attack to the institution and, therefore, ignores that an attack could spread to another institution from a given node and appear on the same institution by attacking another node later. Working with the global \textit{Collaboration Network} would provide more accurate insights about the vulnerability of research institutions as a whole. The current tool would be capable of doing so with small effort on development.

The search engine used for this work does not index all the papers, nor all emails contained in a paper; thus, the network representation is an approximation. Scraping on arXiv \cite{arxiv}, medRxiv \cite{medrxiv}, bioRxiv  \cite{biorxiv} and other similar websites could provide more accurate information. Furthermore, utilizing piracy sites for paper downloads could make easier to target a research center more efficiently. Using these sites to build a global graph and run simulations on it would be outstanding.

It should be noted that some companies also dedicate efforts to publish papers which are indexed by the used search engine; therefore, if they have sufficient workers publishing papers, this strategy could also be useful for that. Most organizations have their email addresses secret, but some of them may be feasible to obtain with the patterns similar to the ones used in this work. An example would be using the LinkedIn \cite{linkedin} social network to target employees, searching for links intelligently (i.e. internet search, presentations from SlideShare \cite{slideshare} or conferences) and building the approximate \textit{Collaboration Network} of an organization.


\section*{Acknowledgements}
We want to thank Iker Gutiérrez-Fandiño for his language review and feedback on the paper.

\section*{Appendix}
This Additional Material includes all figures not presented in the paper \textit{A Vulnerability Study on Academic Collaboration Networks Based on Network Dynamics} that are from deusto.es, upf.edu, ulisboa.pt and urv.cat. However, we wanted to include COVID-19 vaccine research centers to our study that are the Nuffield Department of Medicine (NDM) from the Oxford University ndm.ox.ac.uk and the Spanish National Center of Biotechnology (CNB-CSIC) cnb.csic.es. Note that these are subdomains of larger institutions and the resulting graphs are smaller than the others. We used a higher threshold for selecting the nodes that are moved to exposed or infected to $T=5$. We maintained $\alpha$, $\beta$ and $\gamma$ parameters unchanged.

\begin{sidewaystable}
\centering
\begin{tabular}{@{}lrrrrr@{}}
\toprule
 & \multicolumn{1}{l}{N. nodes} & \multicolumn{1}{l}{N. edges} & \multicolumn{1}{l}{Avg. clustering coeff.} & \multicolumn{1}{l}{Avg. shortest path} & \multicolumn{1}{l}{Deg. assortativity coeff.} \\ \midrule
deusto.es & 96 & 639 & 0.1881 & 2.8070 & 0.4631 \\
ulisboa.pt & 214 & 3,505 & 0.0903 & 2.4679 & 0.4826 \\
upf.edu & 150 & 1,454 & 0.0450 & 2.9428 & 0.5377 \\
urv.cat & 369 & 3,702 & 0.0077 & 4.2027 & 0.5944 \\
ndm.ox.ac.uk & 29 & 107 & 0.1657 & 2.1330 & 0.0101 \\
cnb.csic.es & 51 & 206 & 0.1497 & 2.9066 & 0.2956 \\ \bottomrule
\end{tabular}
\caption{Collaboration Network graph metrics.}
\label{tab:university-metrics-large}
\end{sidewaystable}

\begin{figure}[H]
\centering
\includegraphics[width=7cm]{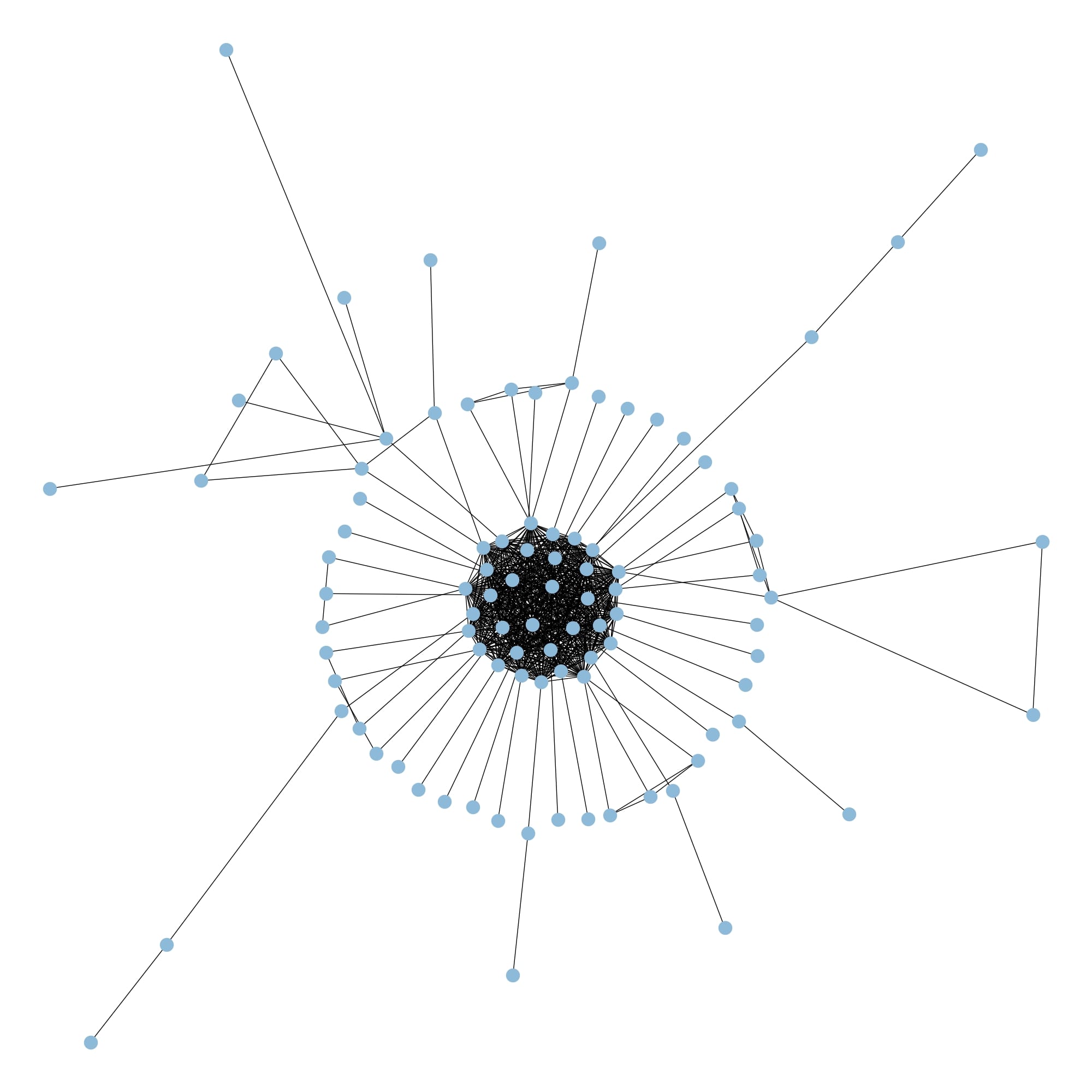}
\caption{Graph plot of deusto.es.}
\end{figure}
\begin{figure}[H]
\centering
\includegraphics[width=7cm]{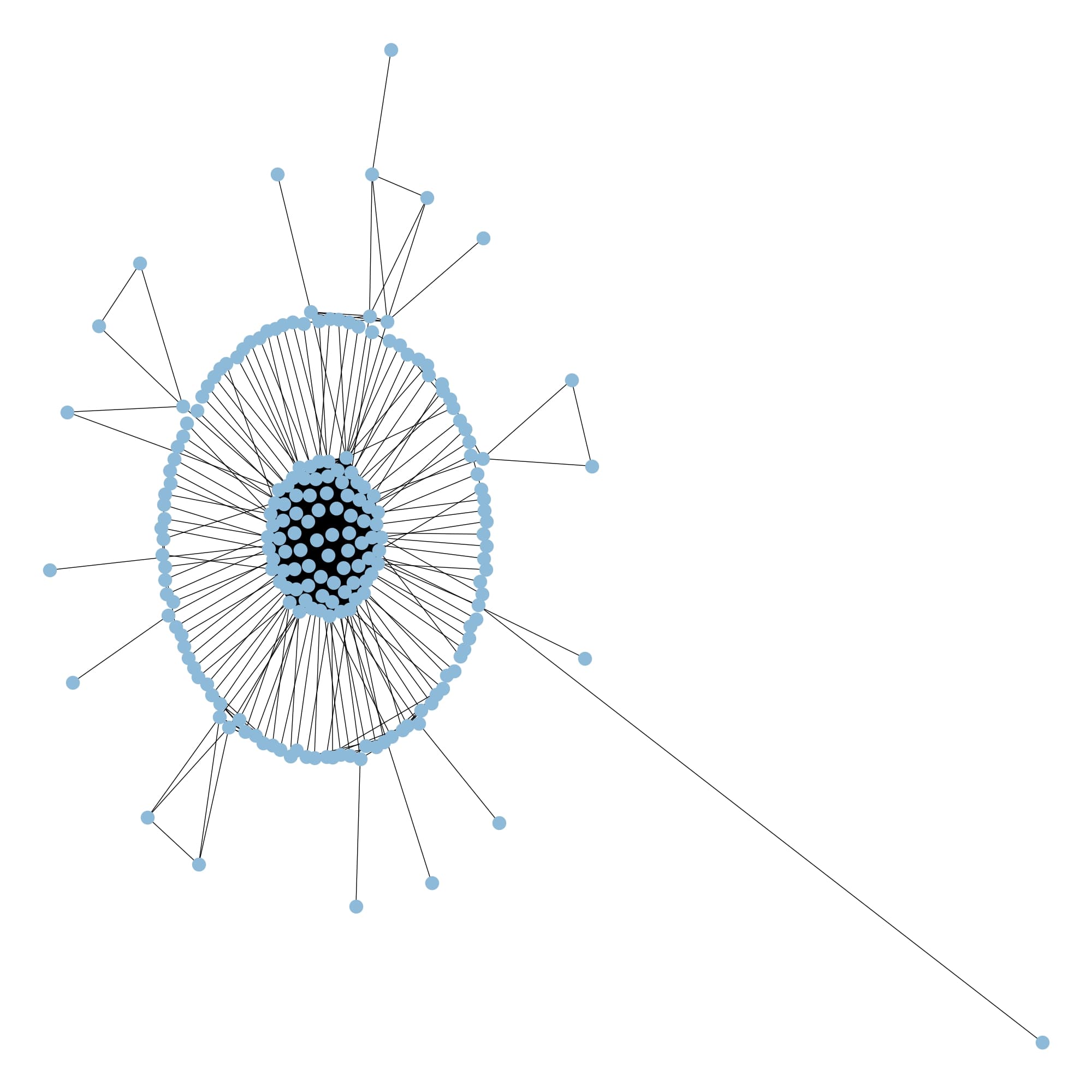}
\caption{Graph plot of ulisboa.pt.}
\end{figure}
\begin{figure}[H]
\centering
\includegraphics[width=7cm]{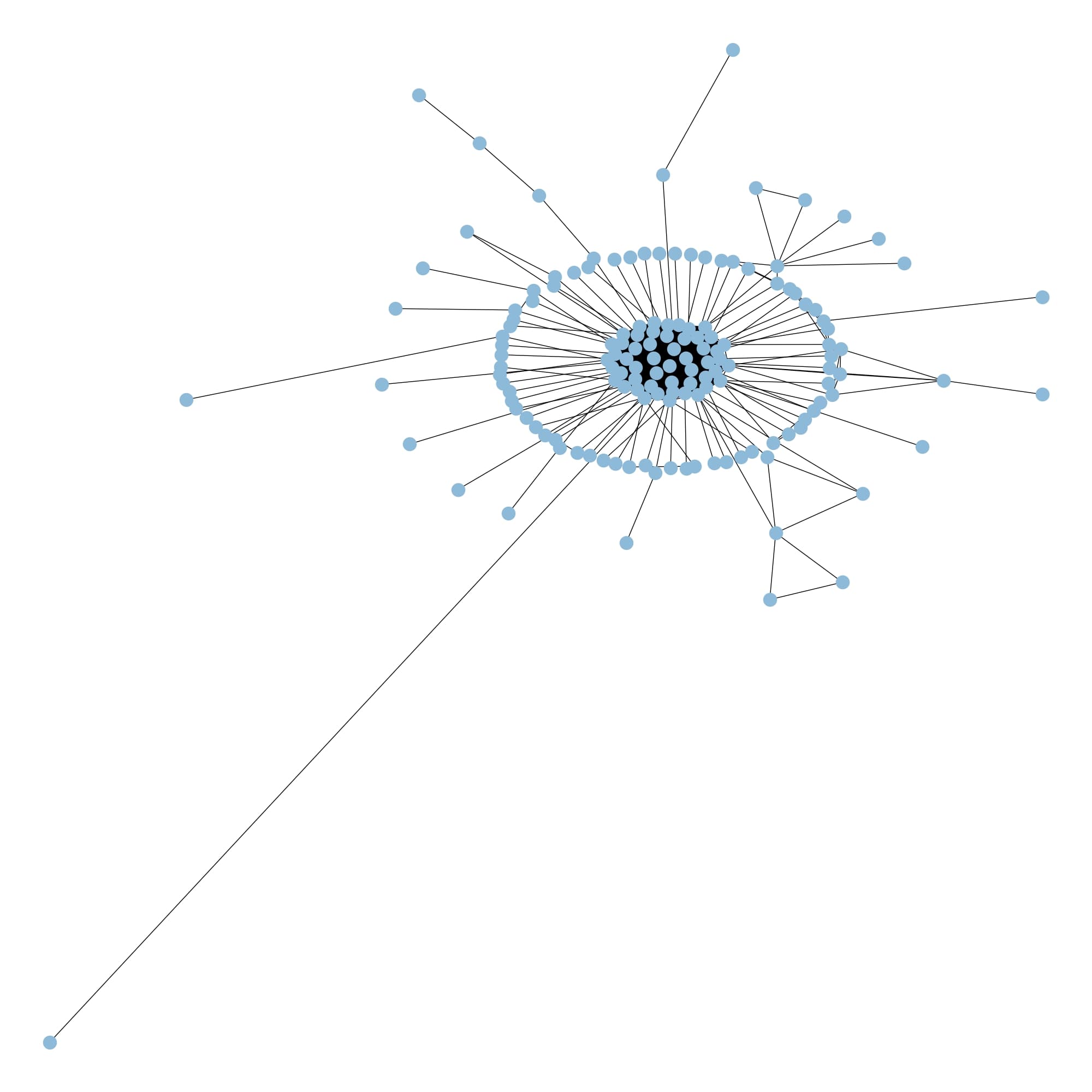}
\caption{Graph plot of upf.edu.}
\end{figure}
\begin{figure}[H]
\centering
\includegraphics[width=7cm]{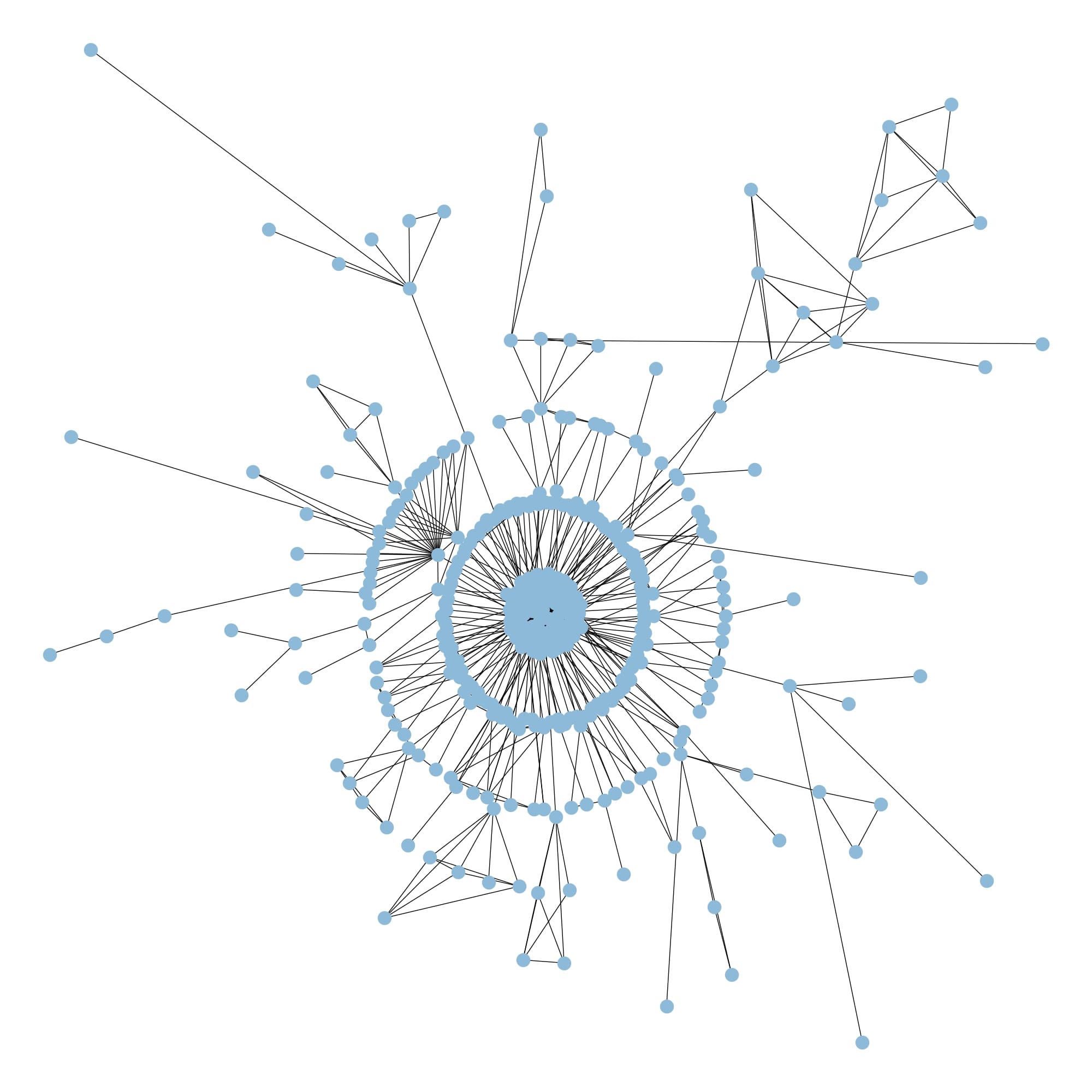}
\caption{Graph plot of urv.cat.}
\end{figure}
\begin{figure}[H]
\centering
\includegraphics[width=7cm]{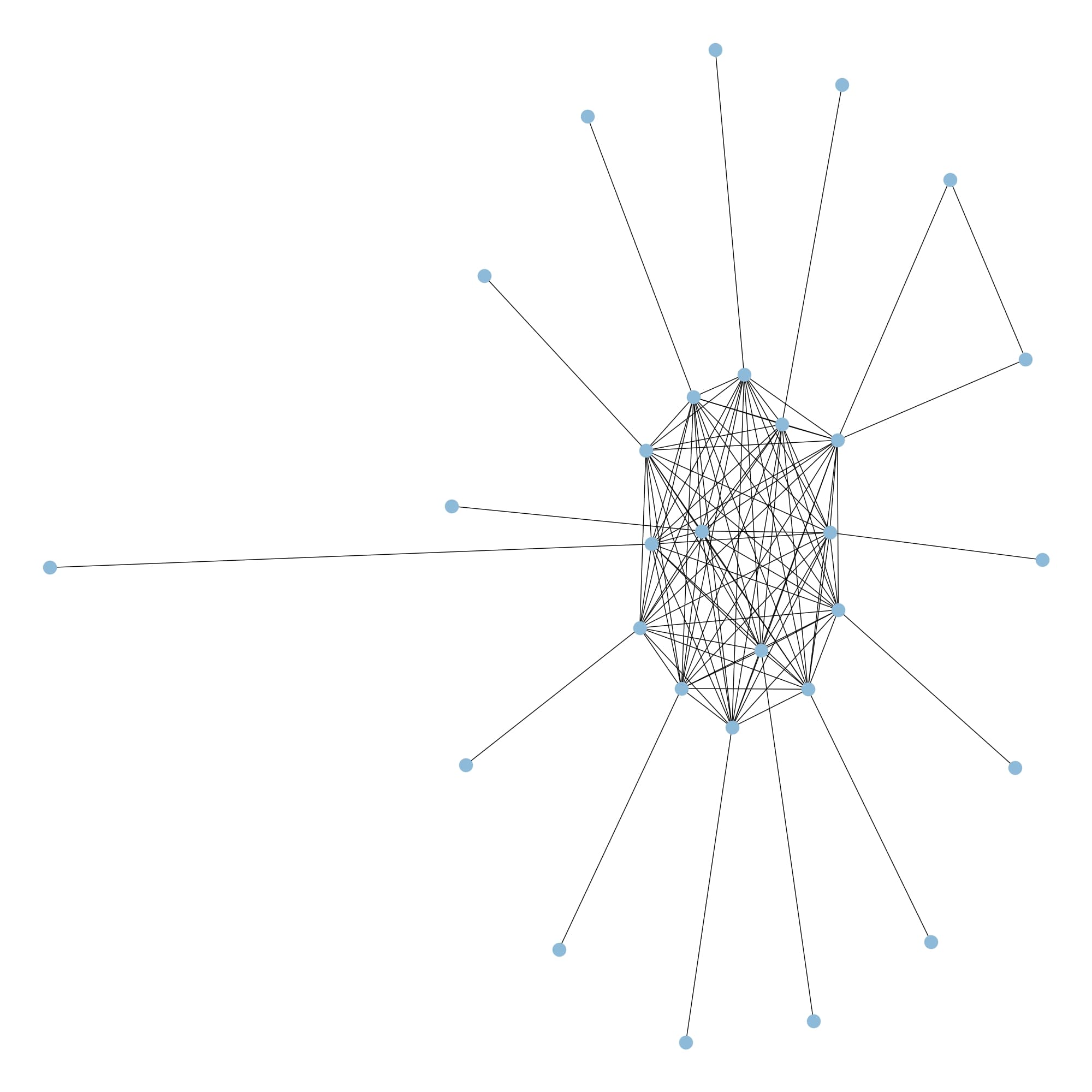}
\caption{Graph plot of ndm.ox.ac.uk.}
\end{figure}
\begin{figure}[H]
\centering
\includegraphics[width=7cm]{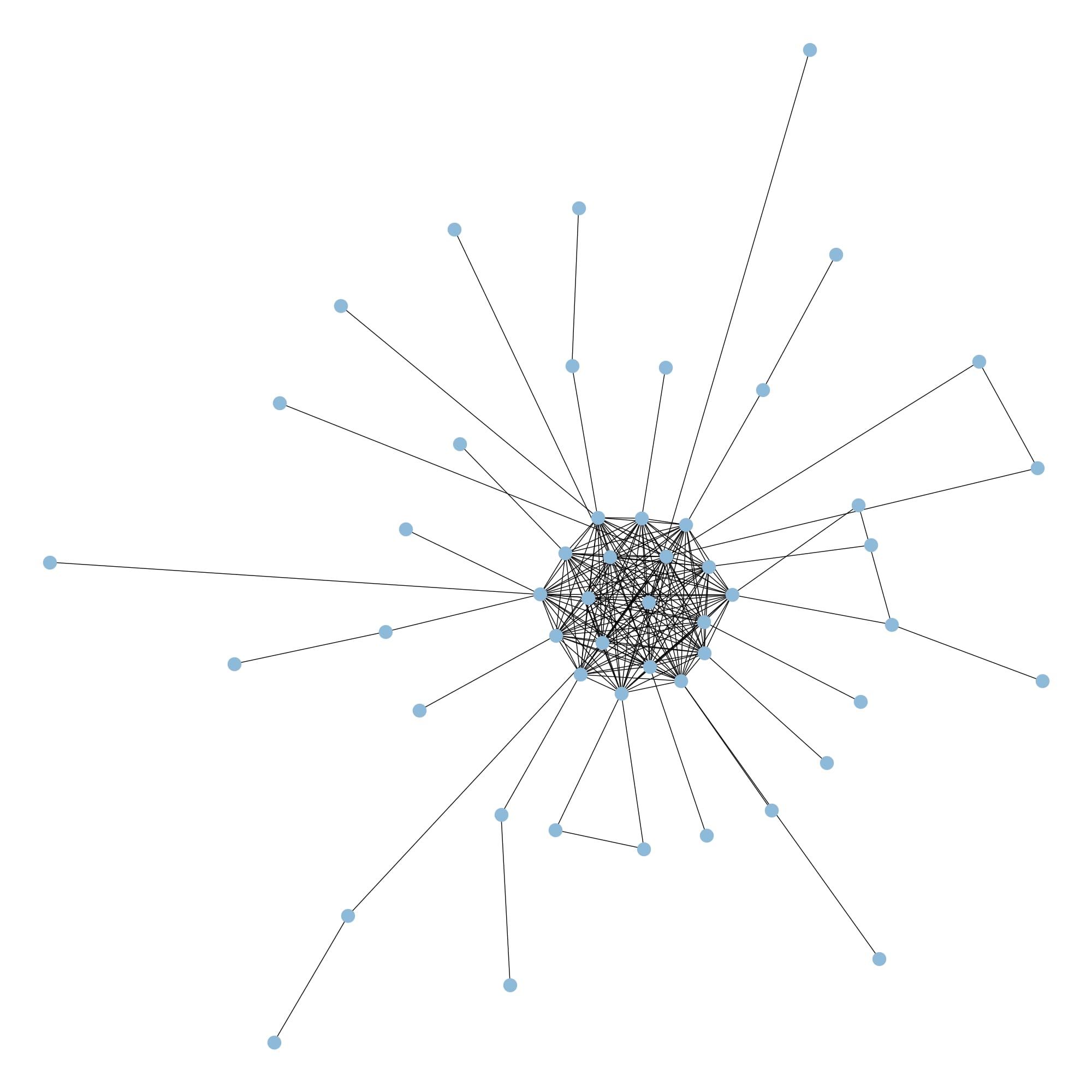}
\caption{Graph plot of cnb.csic.es.}
\end{figure}

\begin{figure}[H]
\centering
\includegraphics[width=7cm]{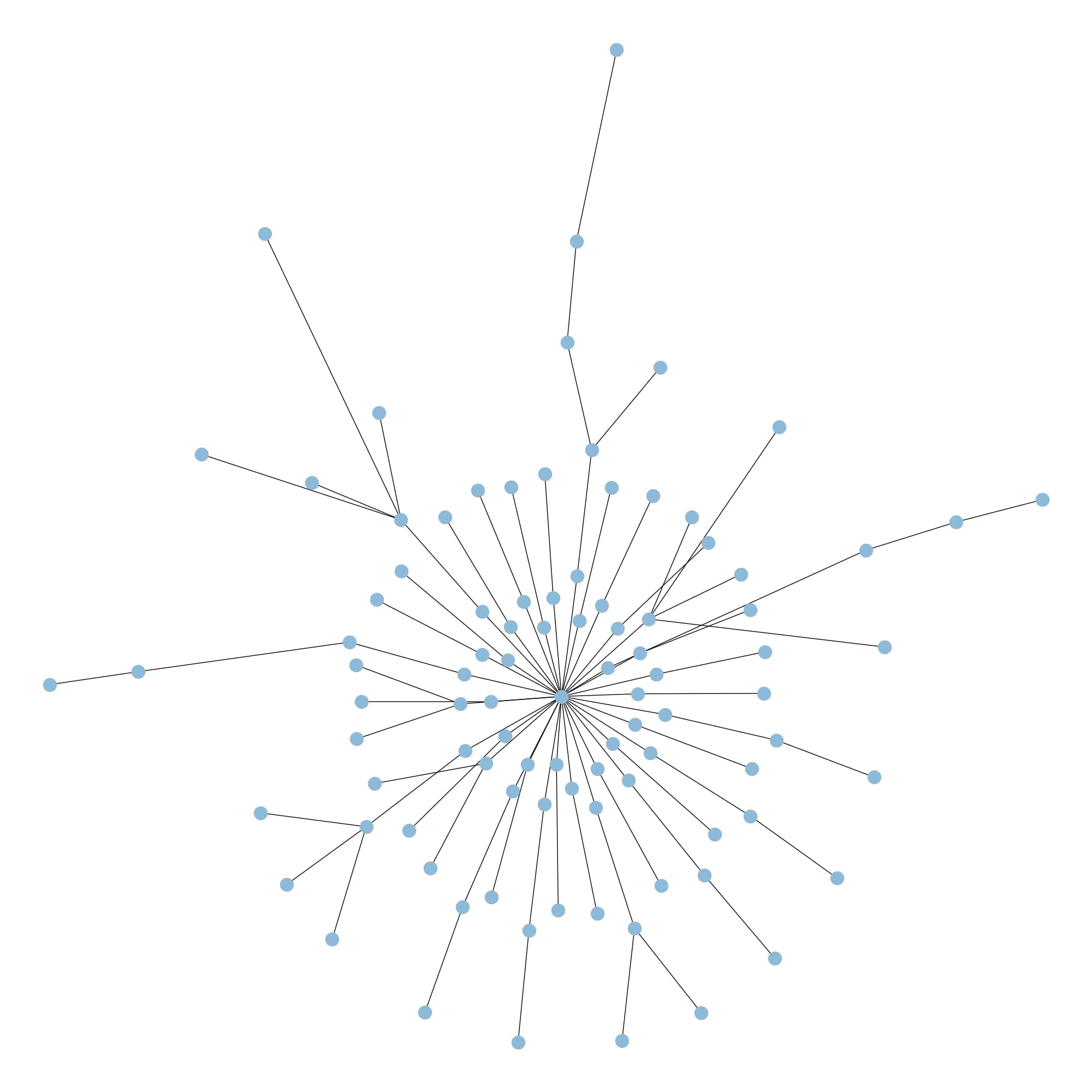}
\caption{MST plot of deusto.es.}
\end{figure}
\begin{figure}[H]
\centering
\includegraphics[width=7cm]{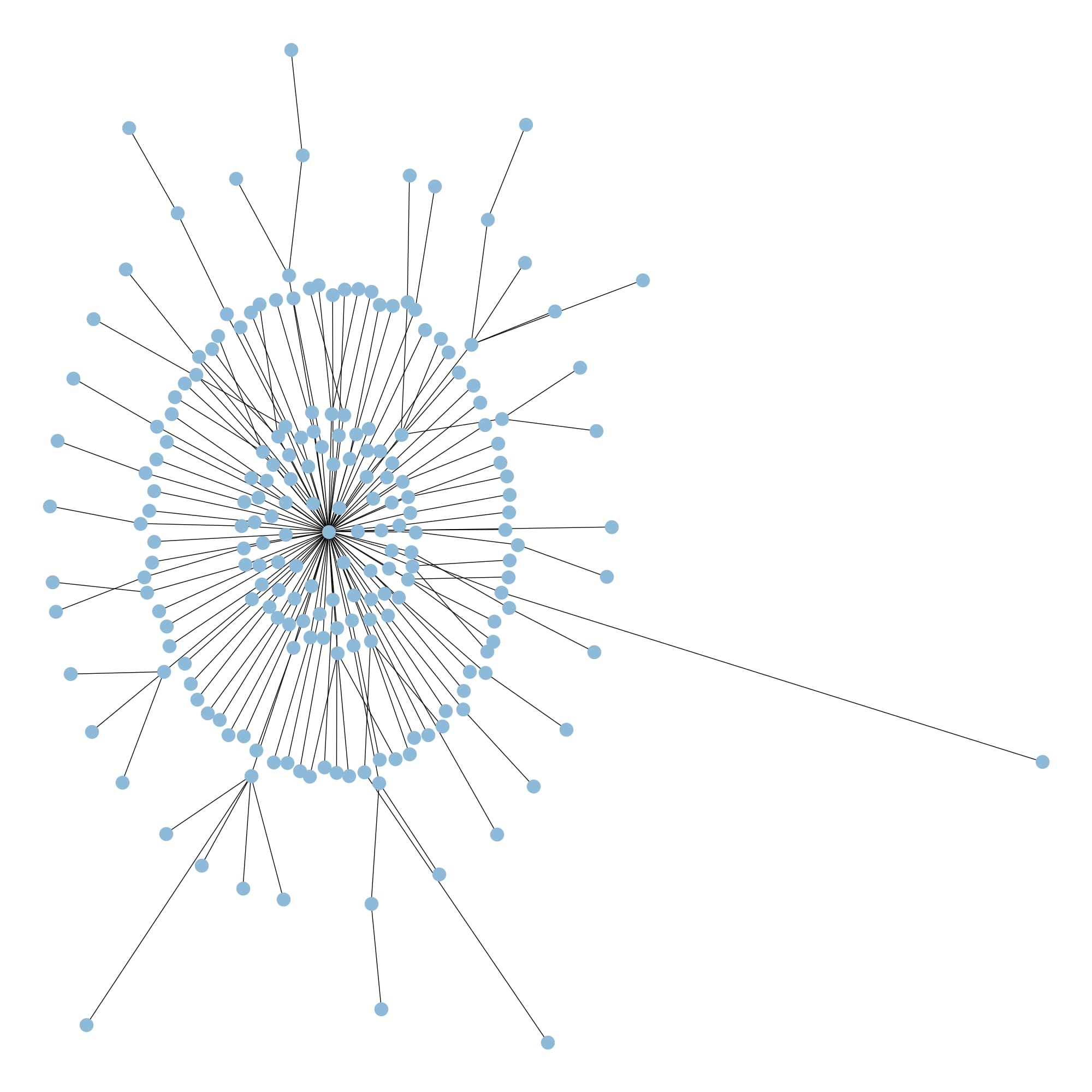}
\caption{MST plot of ulisboa.pt.}
\end{figure}
\begin{figure}[H]
\centering
\includegraphics[width=7cm]{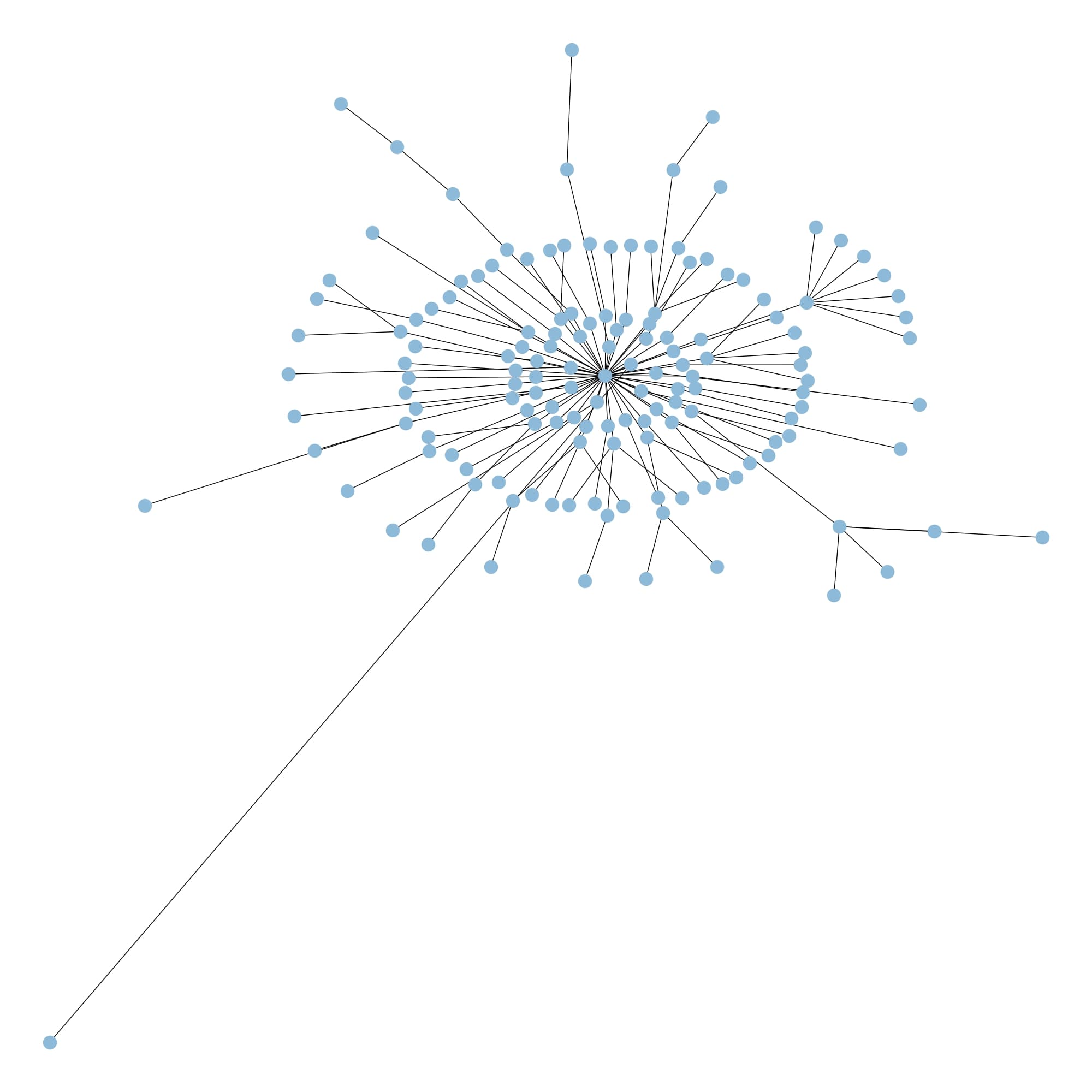}
\caption{MST plot of upf.edu.}
\end{figure}
\begin{figure}[H]
\centering
\includegraphics[width=7cm]{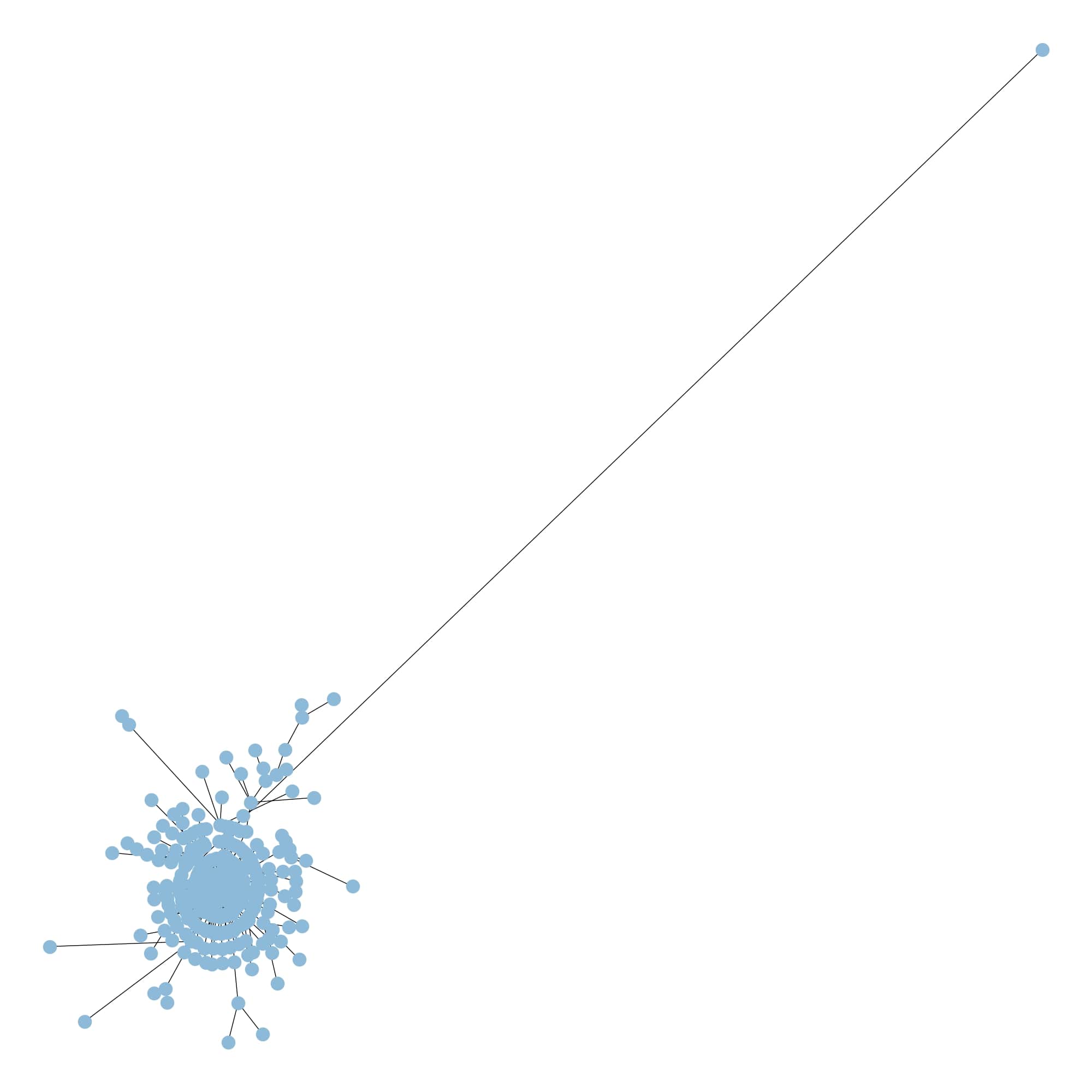}
\caption{MST plot of urv.cat.}
\end{figure}
\begin{figure}[H]
\centering
\includegraphics[width=7cm]{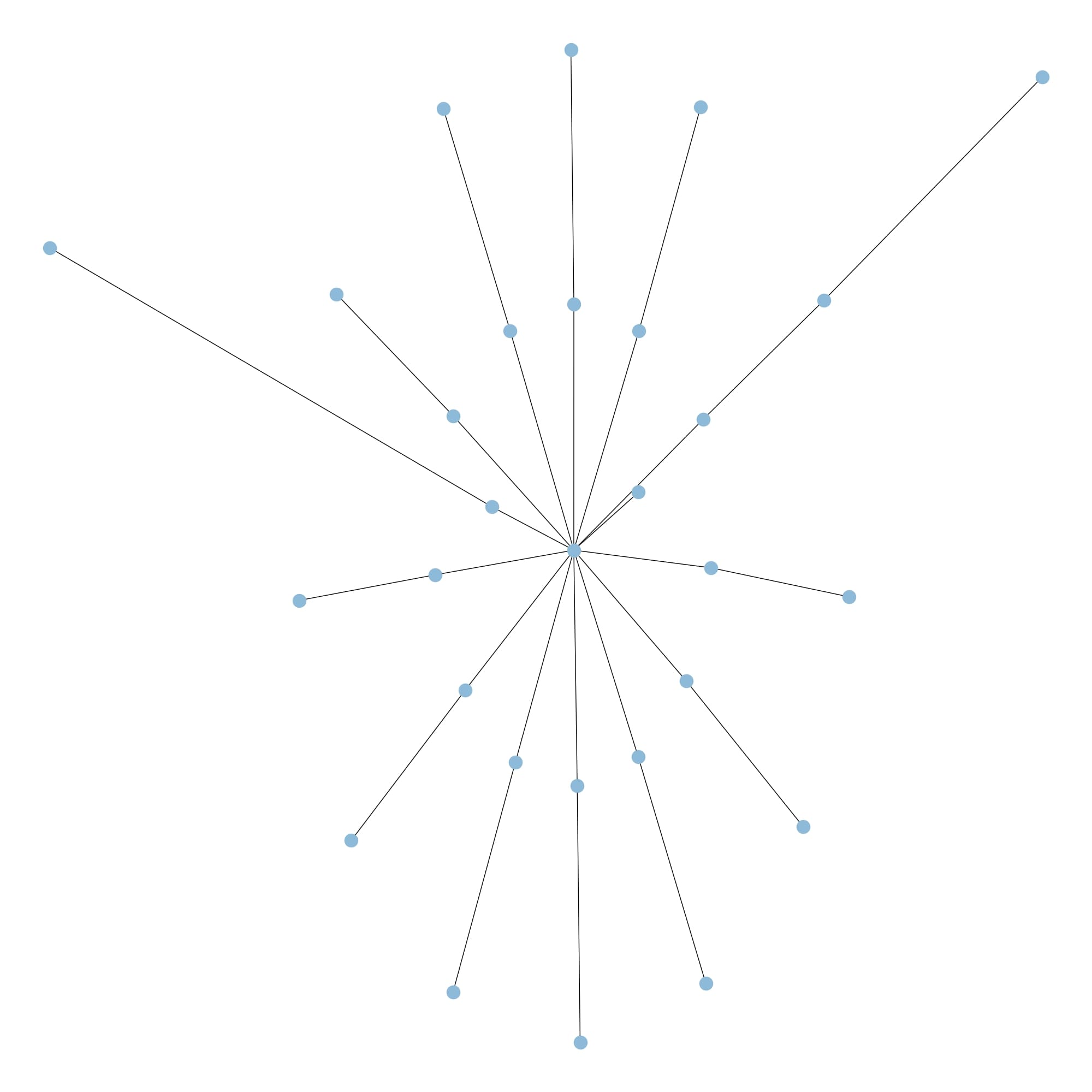}
\caption{MST plot of ndm.ox.ac.uk.}
\end{figure}
\begin{figure}[H]
\centering
\includegraphics[width=7cm]{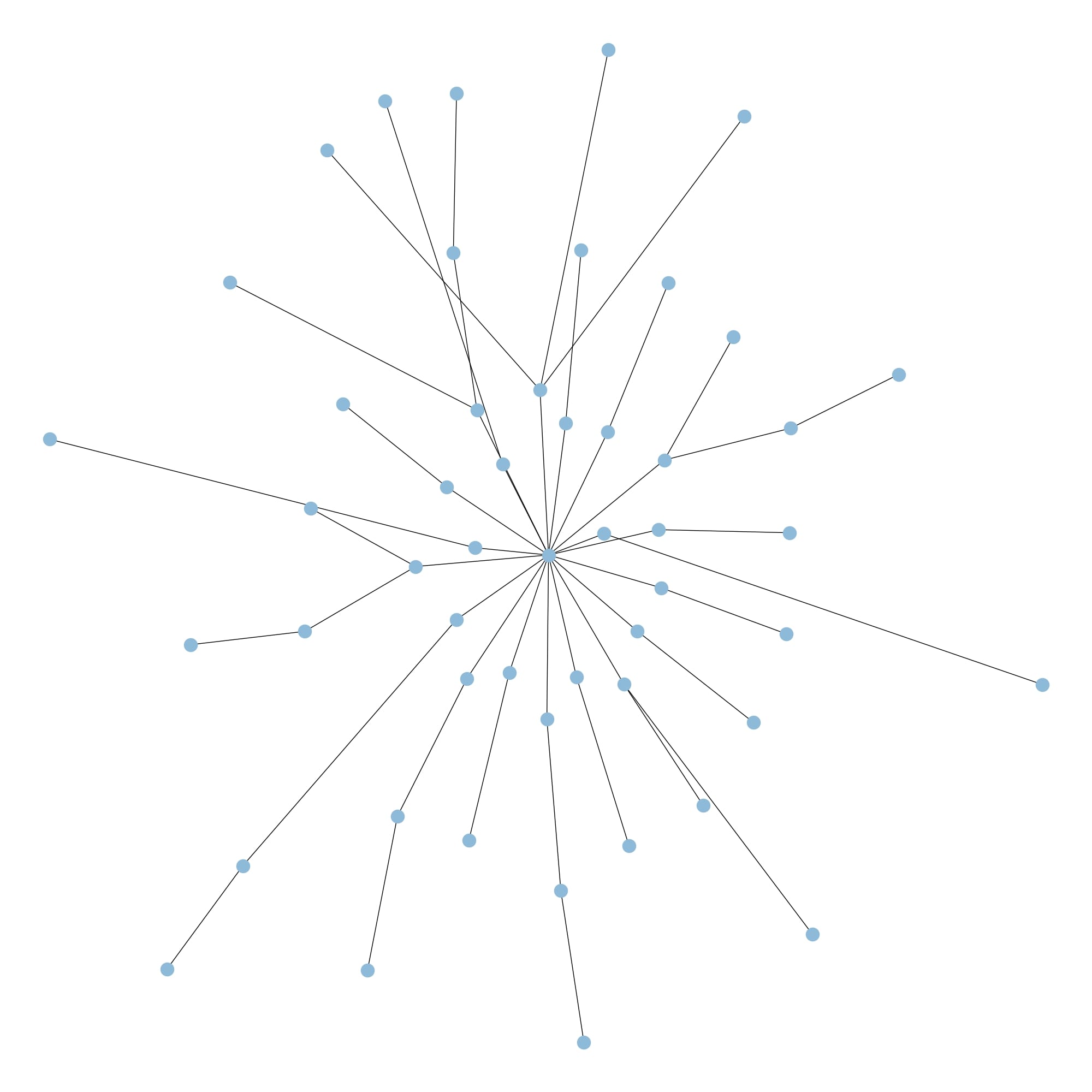}
\caption{MST plot of cnb.csic.es.}
\end{figure}

\begin{figure}[H]
\centering
\includegraphics[width=7cm]{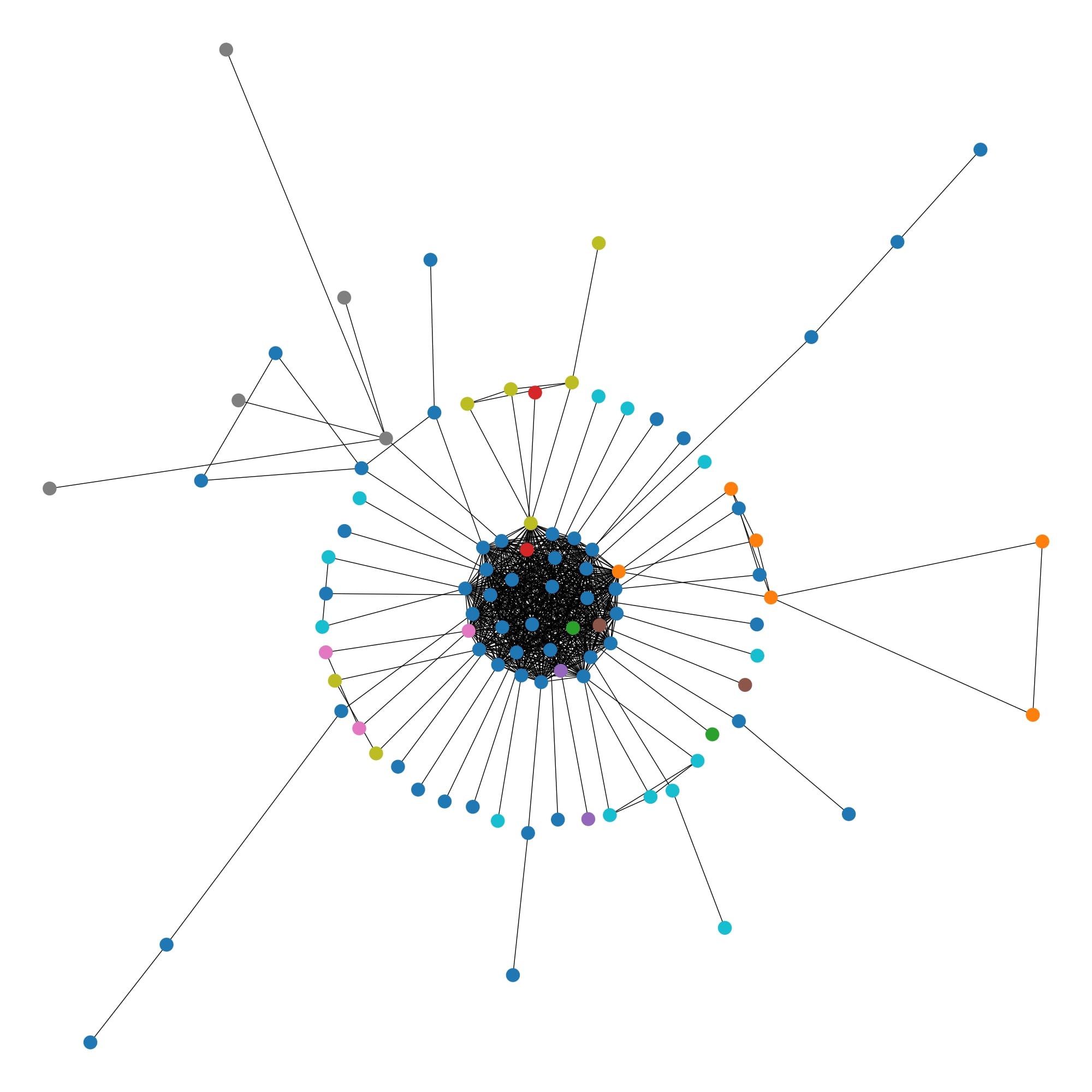}
\caption{Community Detection plot of deusto.es.}
\end{figure}
\begin{figure}[H]
\centering
\includegraphics[width=7cm]{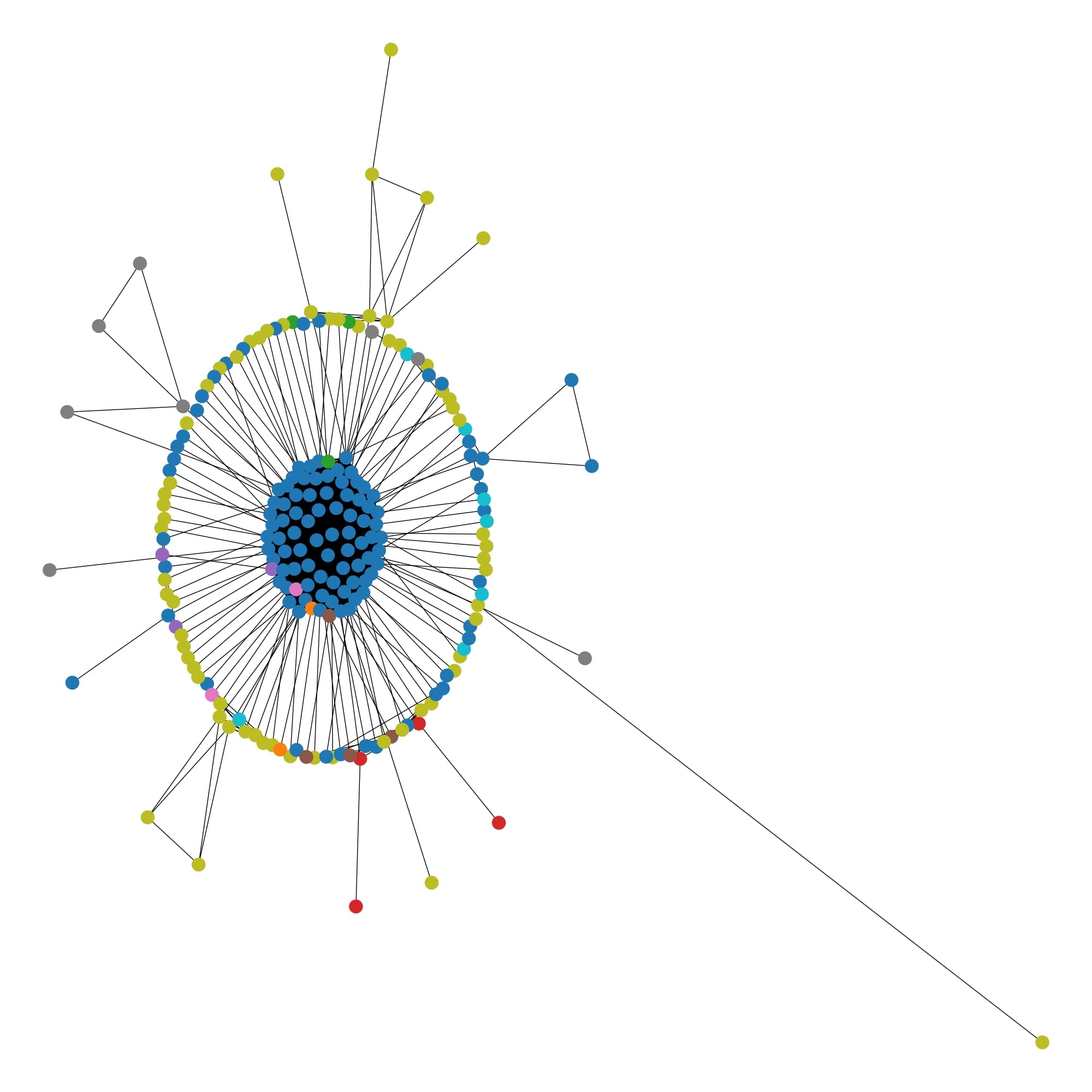}
\caption{Community Detection plot of ulisboa.pt.}
\end{figure}
\begin{figure}[H]
\centering
\includegraphics[width=7cm]{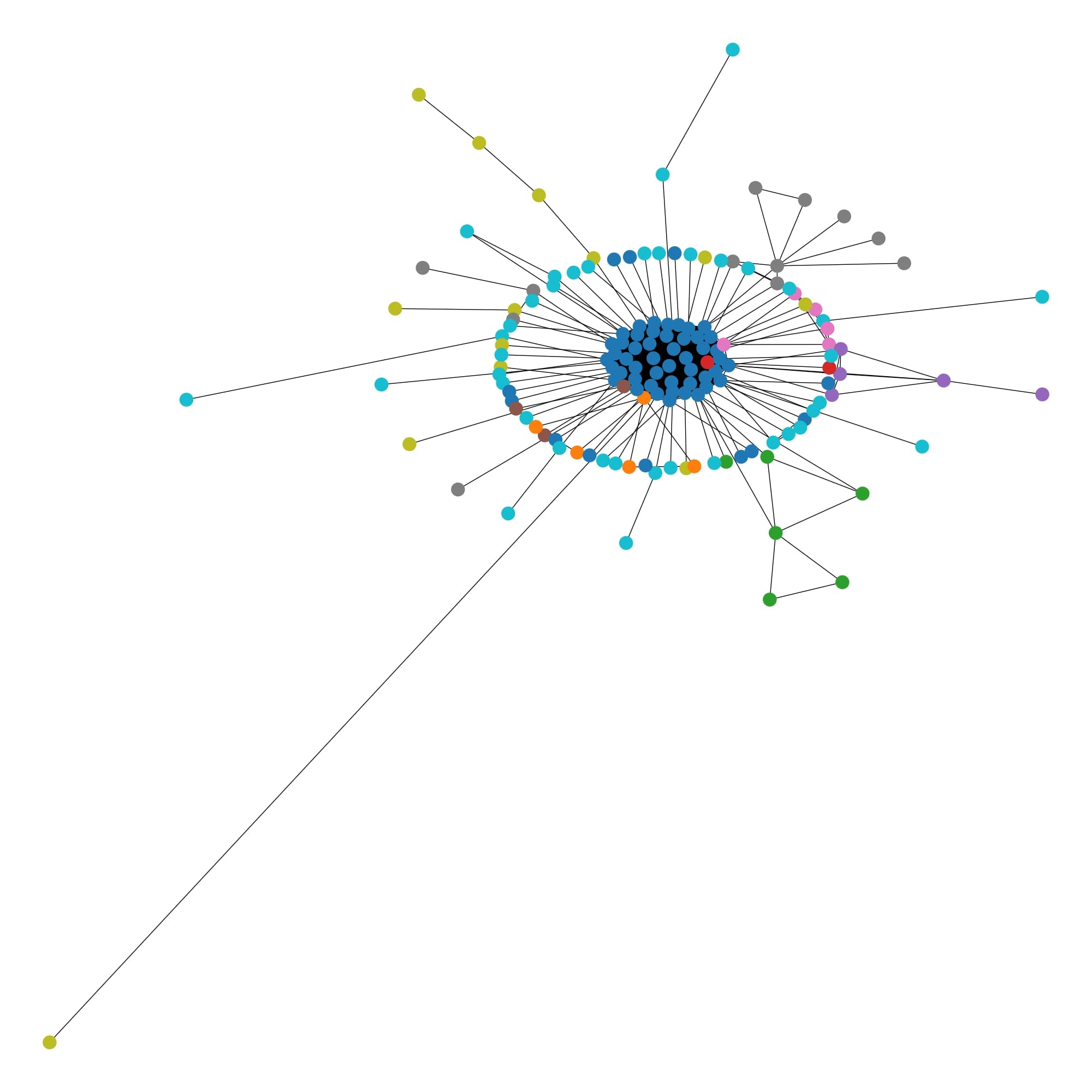}
\caption{Community Detection plot of upf.edu.}
\end{figure}
\begin{figure}[H]
\centering
\includegraphics[width=7cm]{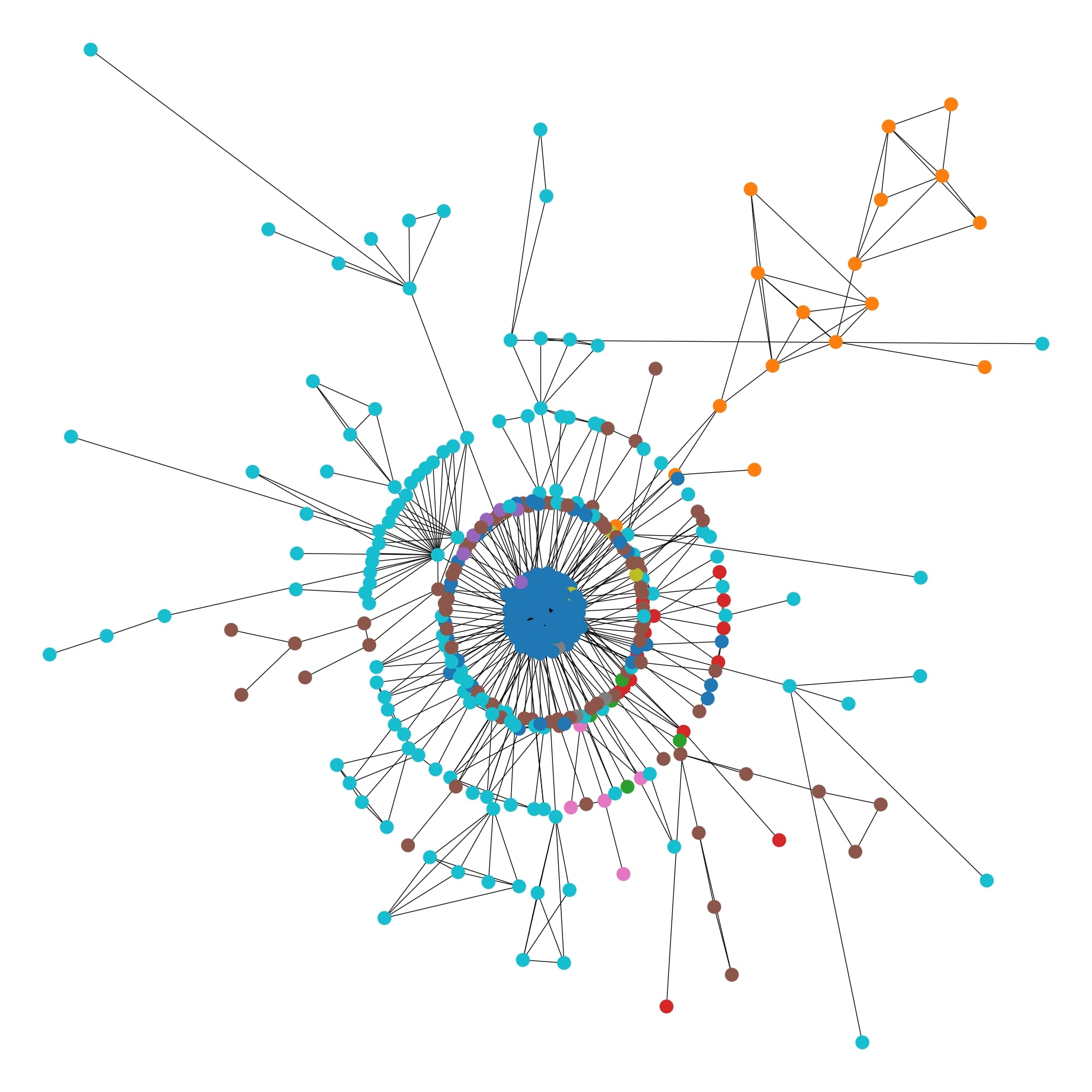}
\caption{Community Detection plot of urv.cat.}
\end{figure}
\begin{figure}[H]
\centering
\includegraphics[width=7cm]{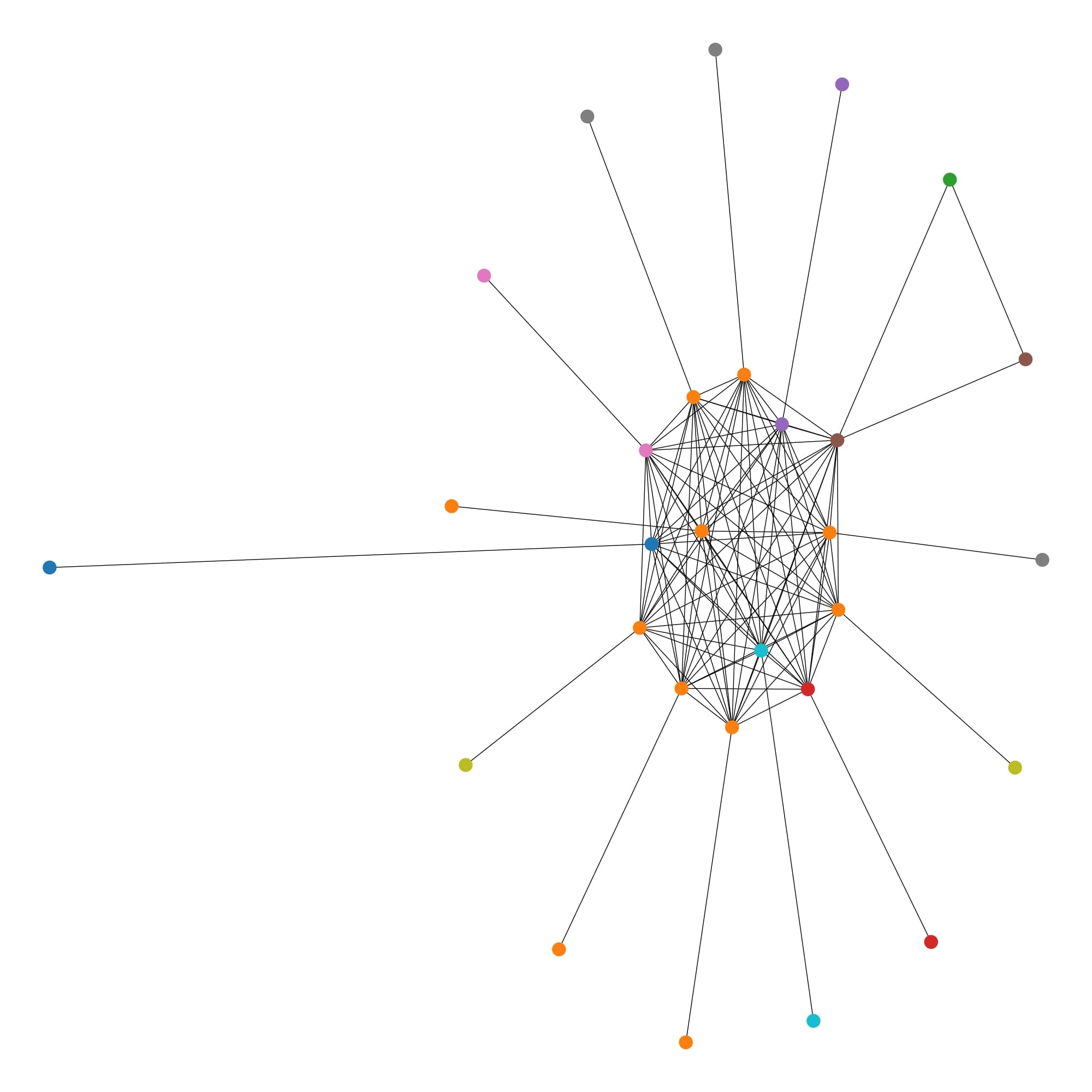}
\caption{Community Detection plot of ndm.ox.ac.uk.}
\end{figure}
\begin{figure}[H]
\centering
\includegraphics[width=7cm]{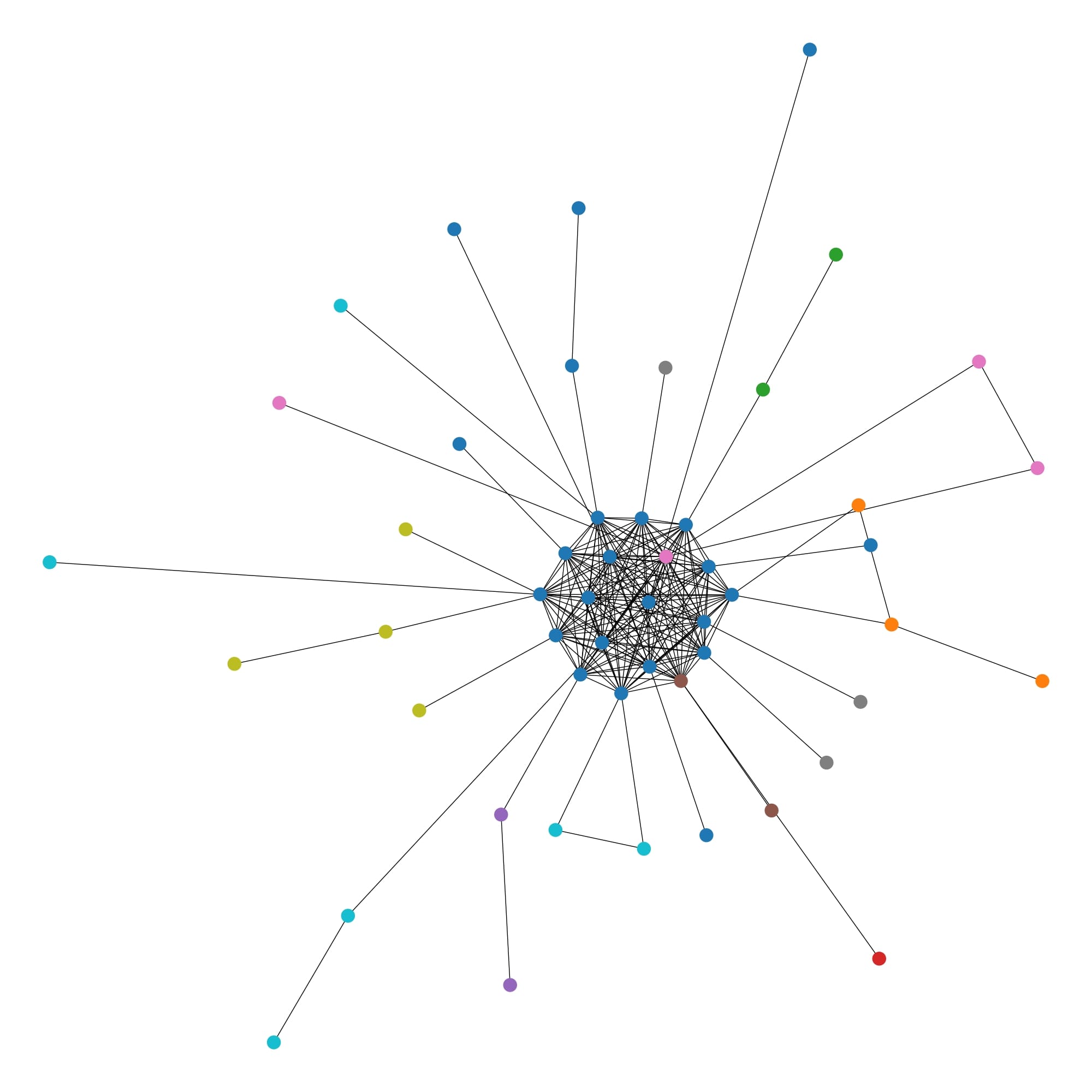}
\caption{Community Detection plot of cnb.csic.es.}
\end{figure}

\begin{figure}[H]
\centering
\includegraphics[width=7cm]{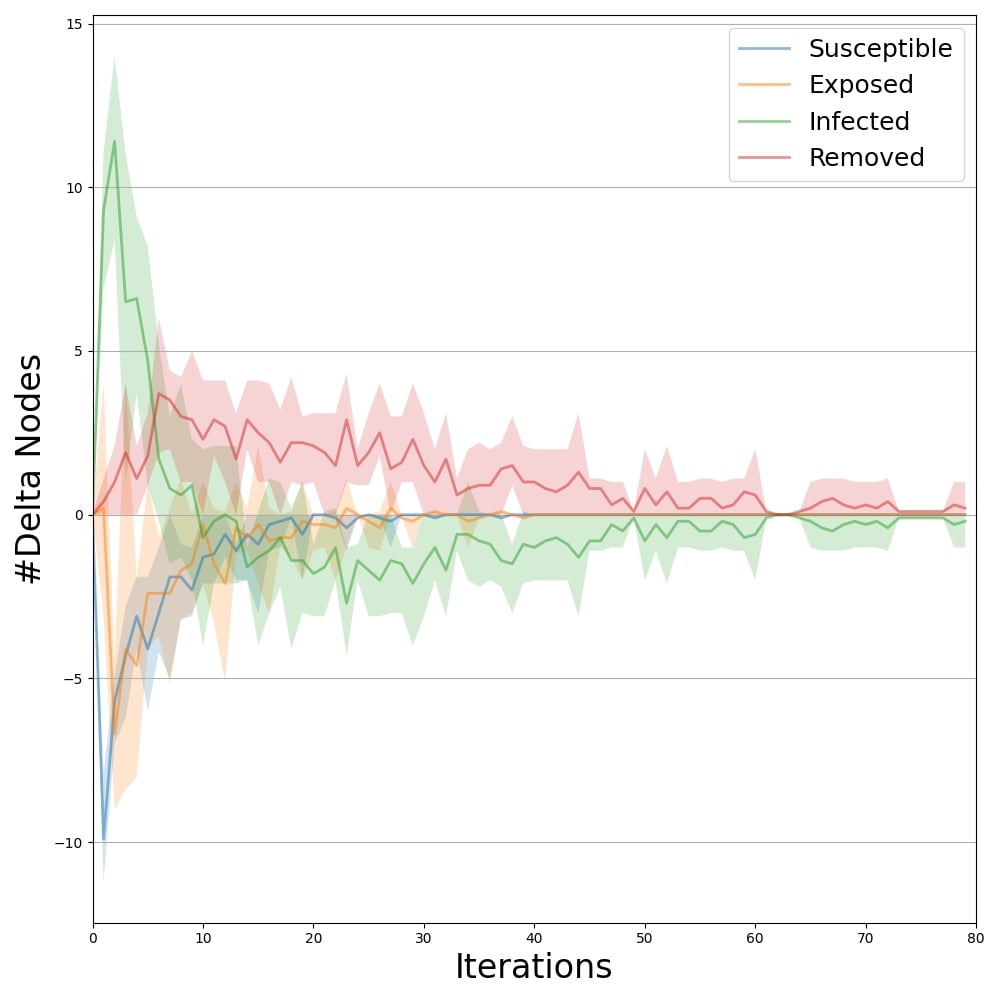}
\caption{Diffusion Prevalence diagram of deusto.es.}
\end{figure}
\begin{figure}[H]
\centering
\includegraphics[width=7cm]{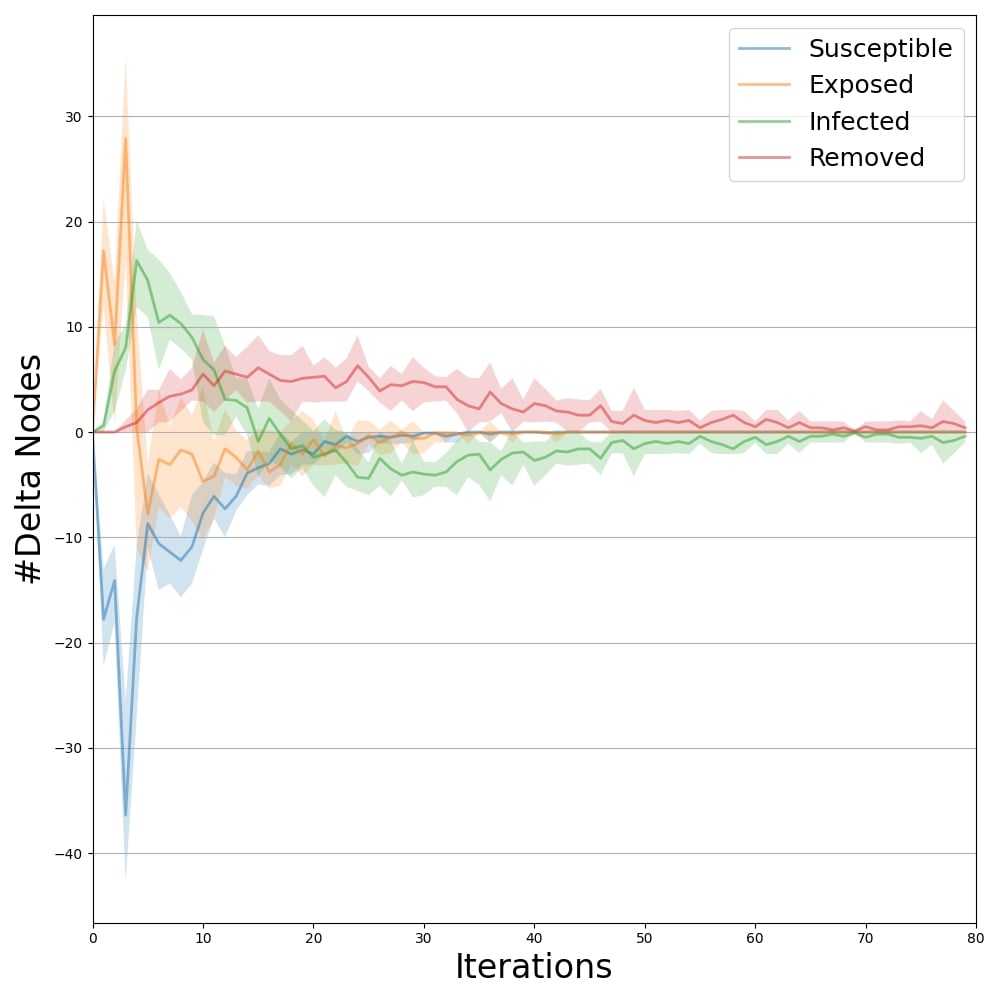}
\caption{Diffusion Prevalence diagram of ulisboa.pt.}
\end{figure}
\begin{figure}[H]
\centering
\includegraphics[width=7cm]{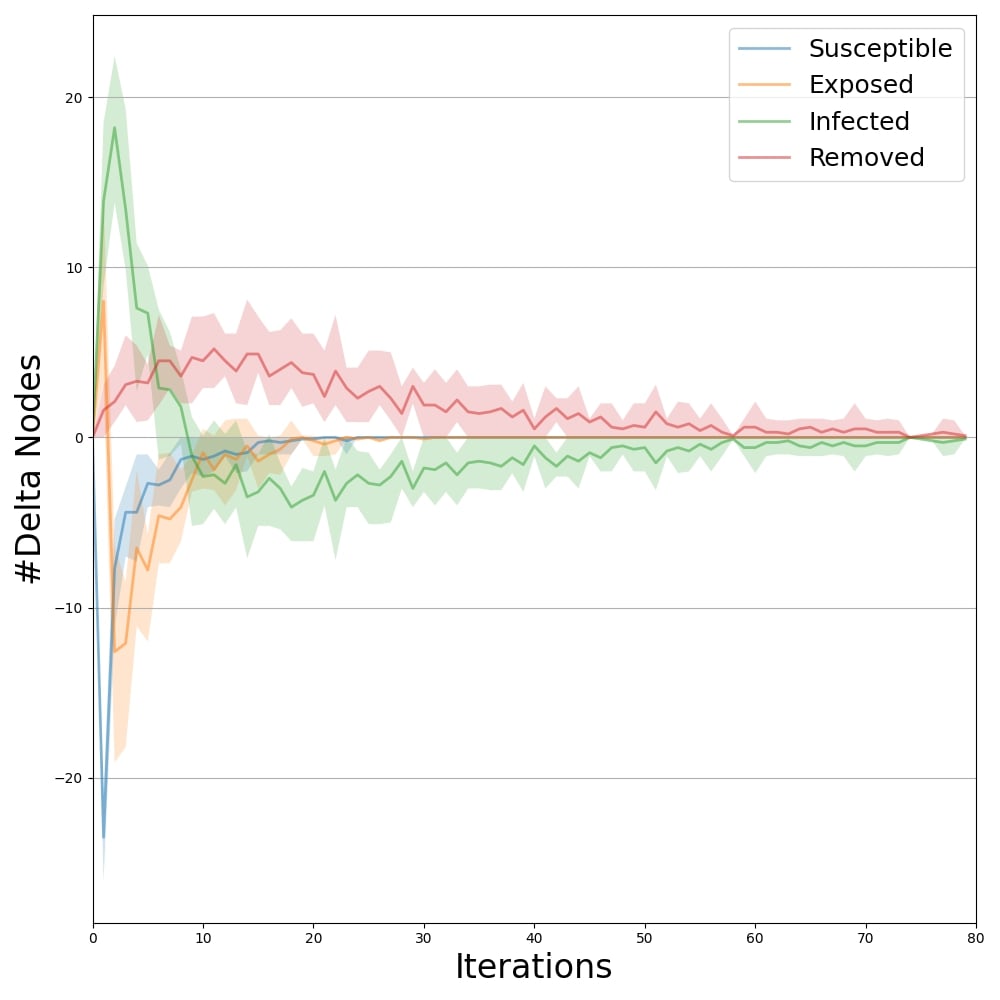}
\caption{Diffusion Prevalence diagram of upf.edu.}
\end{figure}
\begin{figure}[H]
\centering
\includegraphics[width=7cm]{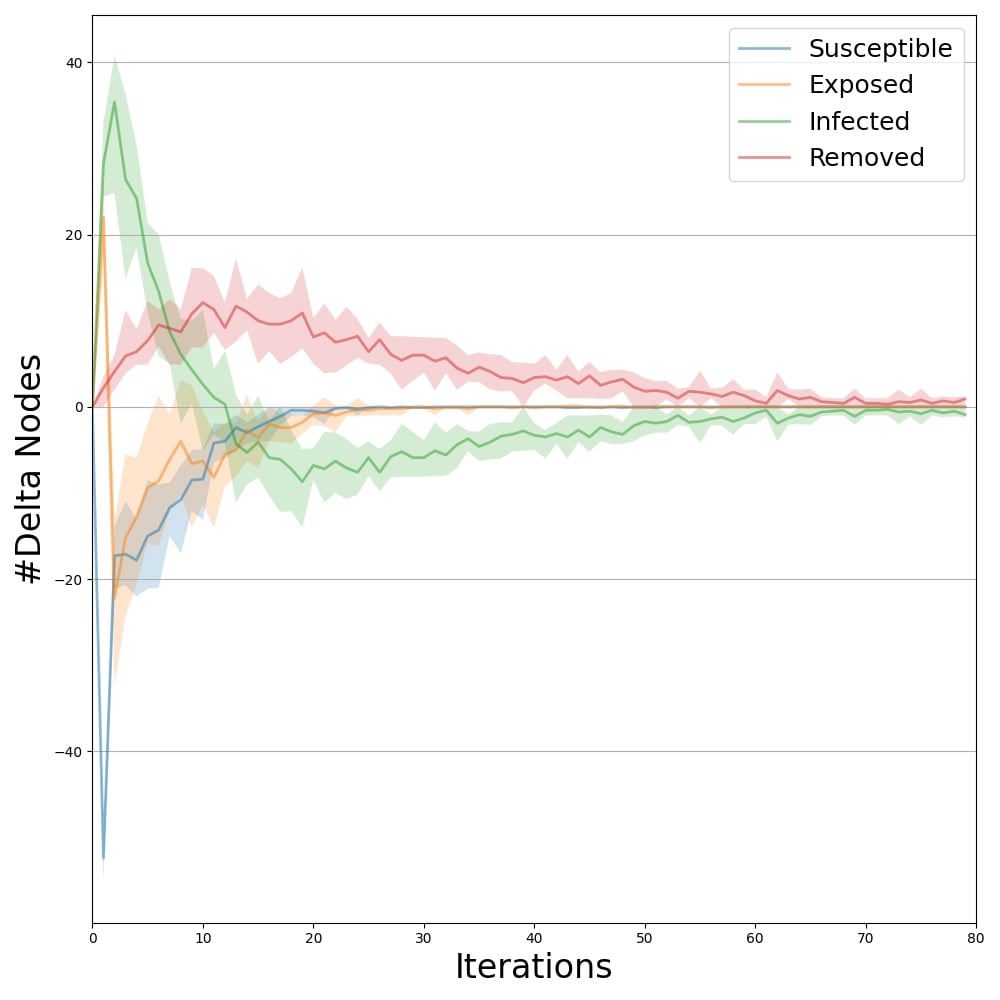}
\caption{Diffusion Prevalence diagram of urv.cat.}
\end{figure}
\begin{figure}[H]
\centering
\includegraphics[width=7cm]{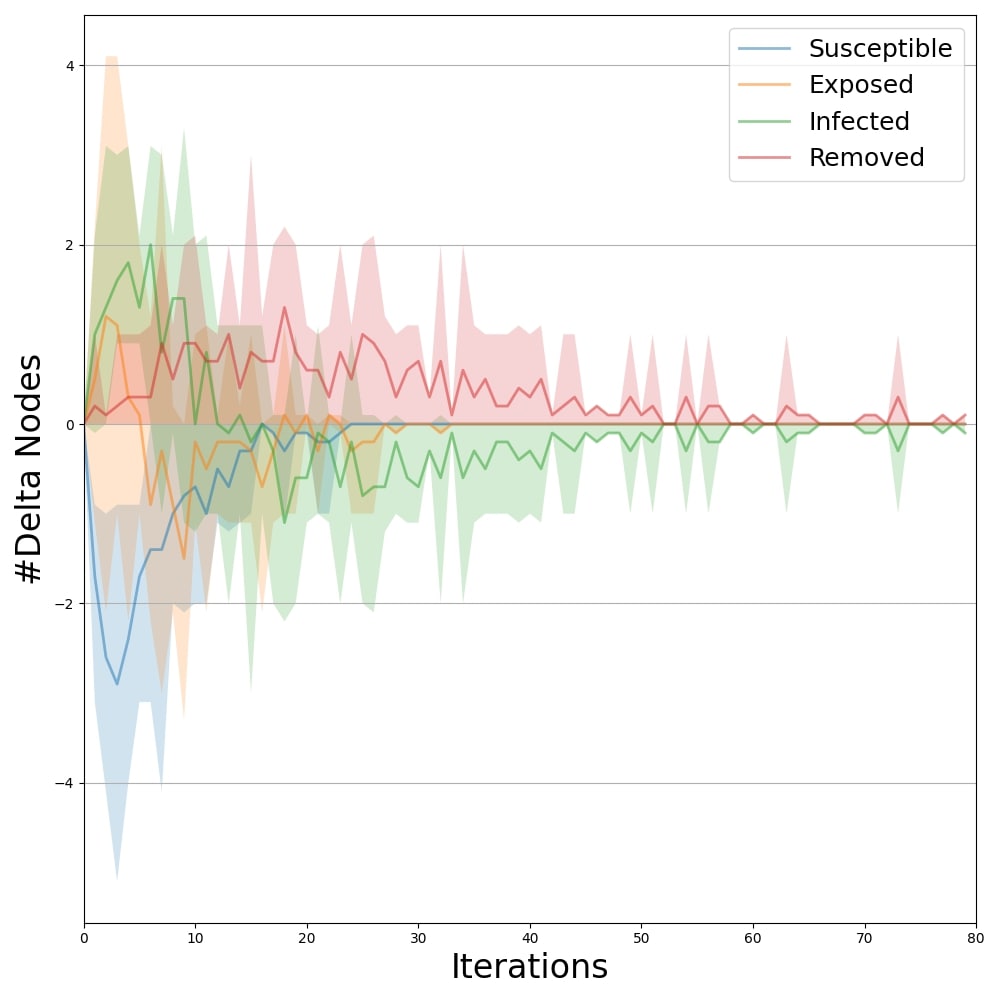}
\caption{Diffusion Prevalence diagram of ndm.ox.ac.uk.}
\end{figure}
\begin{figure}[H]
\centering
\includegraphics[width=7cm]{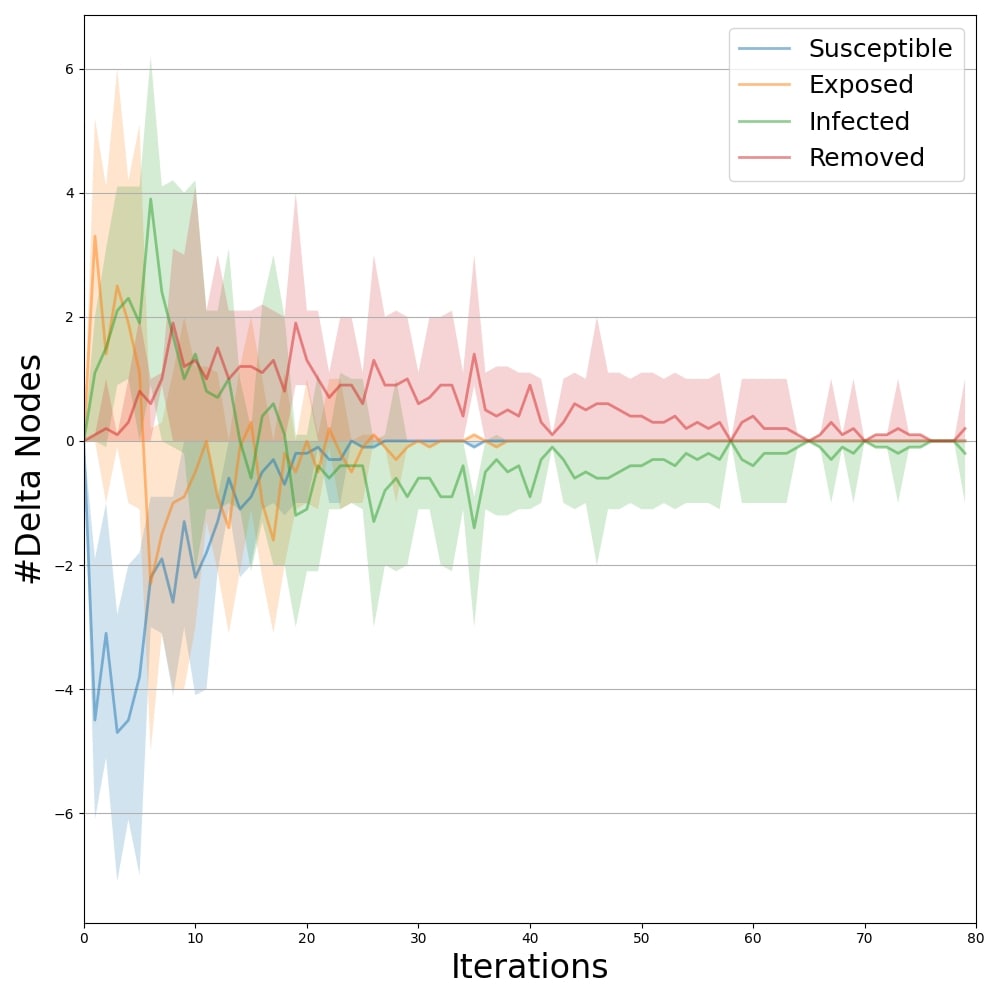}
\caption{Diffusion Prevalence diagram of cnb.csic.es.}
\end{figure}

\begin{figure}[H]
\centering
\includegraphics[width=7cm]{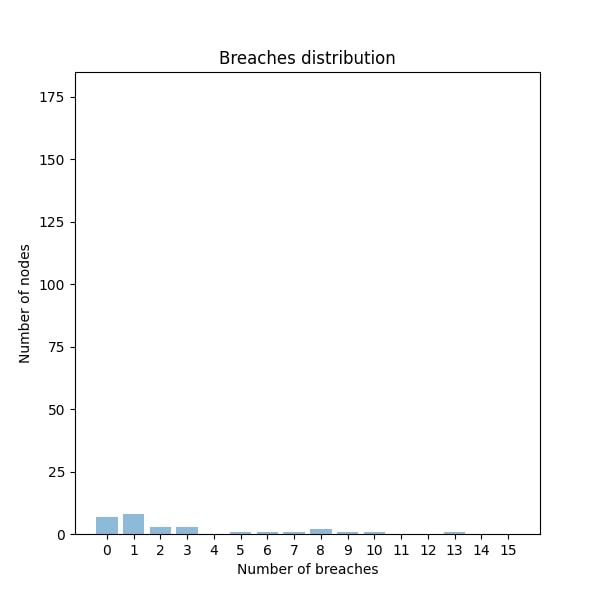}
\caption{Breach distribution bar plot of ndm.ox.ac.uk.}
\end{figure}
\begin{figure}[H]
\centering
\includegraphics[width=7cm]{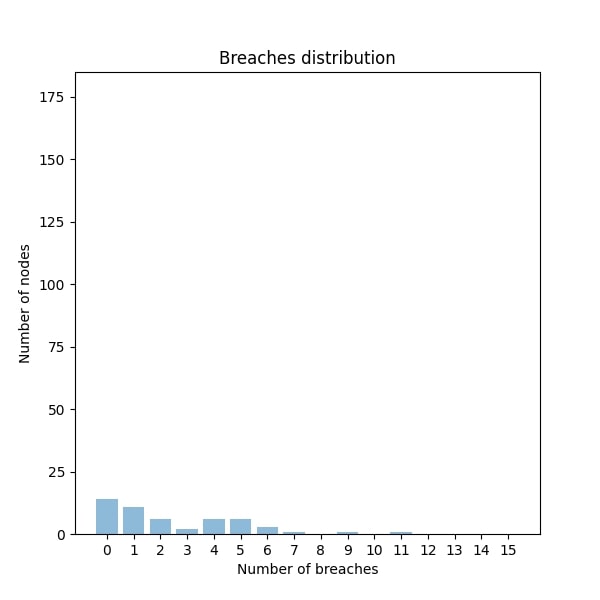}
\caption{Breach distribution bar plot of cnb.csic.es.}
\end{figure}

\begin{figure}[H]
\centering
\includegraphics[width=7cm]{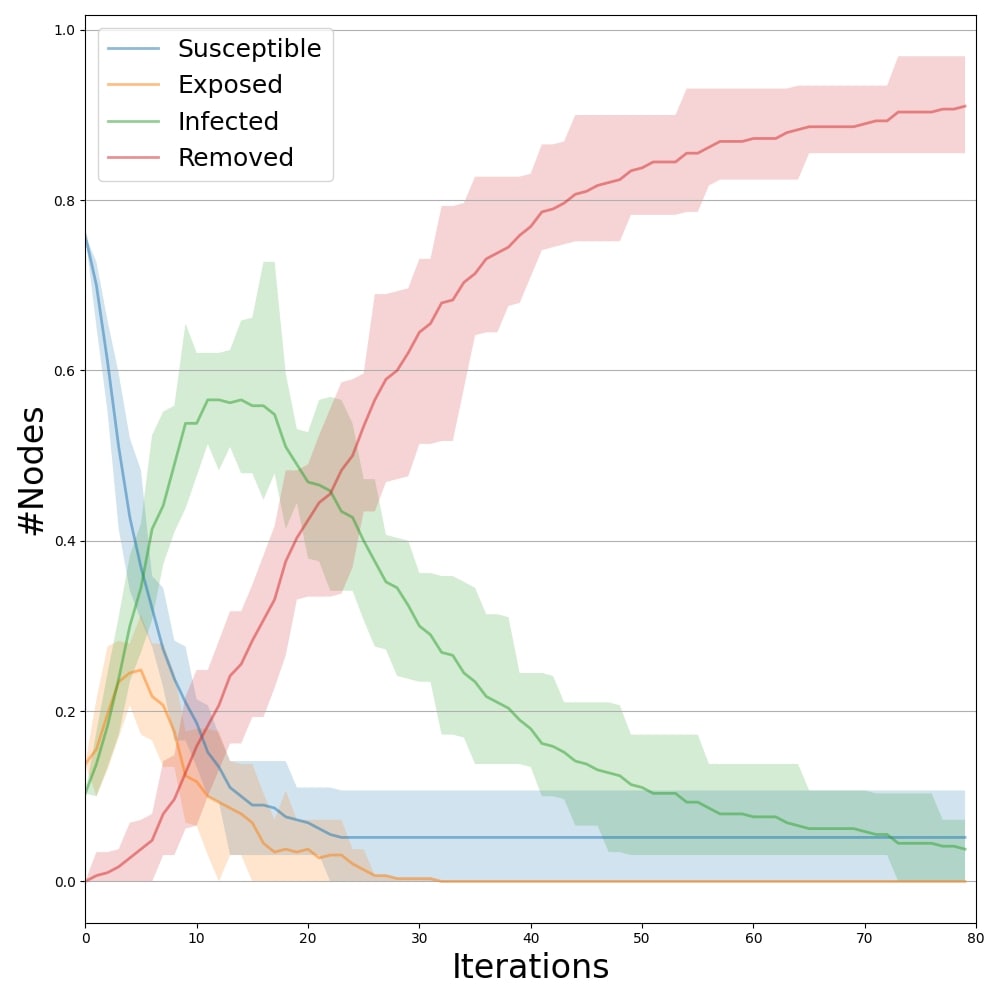}
\caption{Diffusion Trend diagram of ndm.ox.ac.uk.}
\end{figure}
\begin{figure}[H]
\centering
\includegraphics[width=7cm]{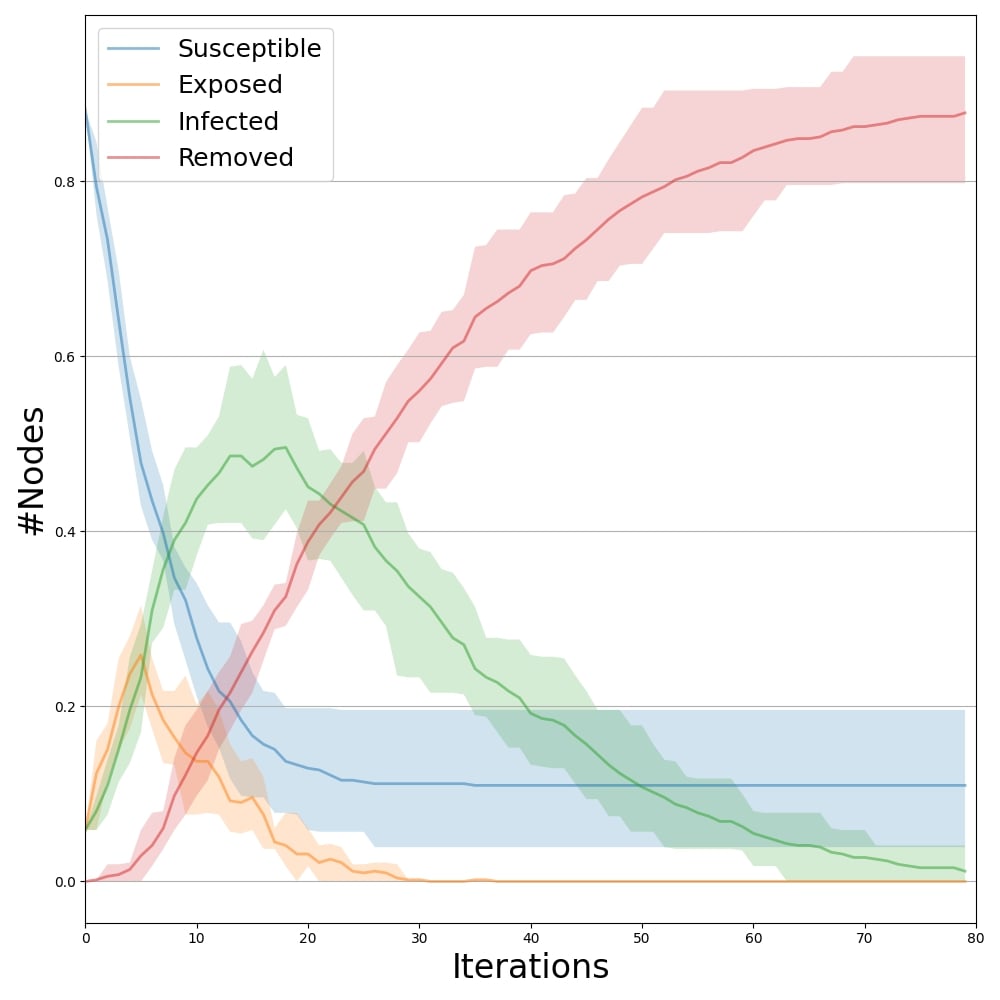}
\caption{Diffusion Trend diagram of cnb.csic.es.}
\end{figure}

\begin{figure}[H]
\centering
\includegraphics[width=7cm]{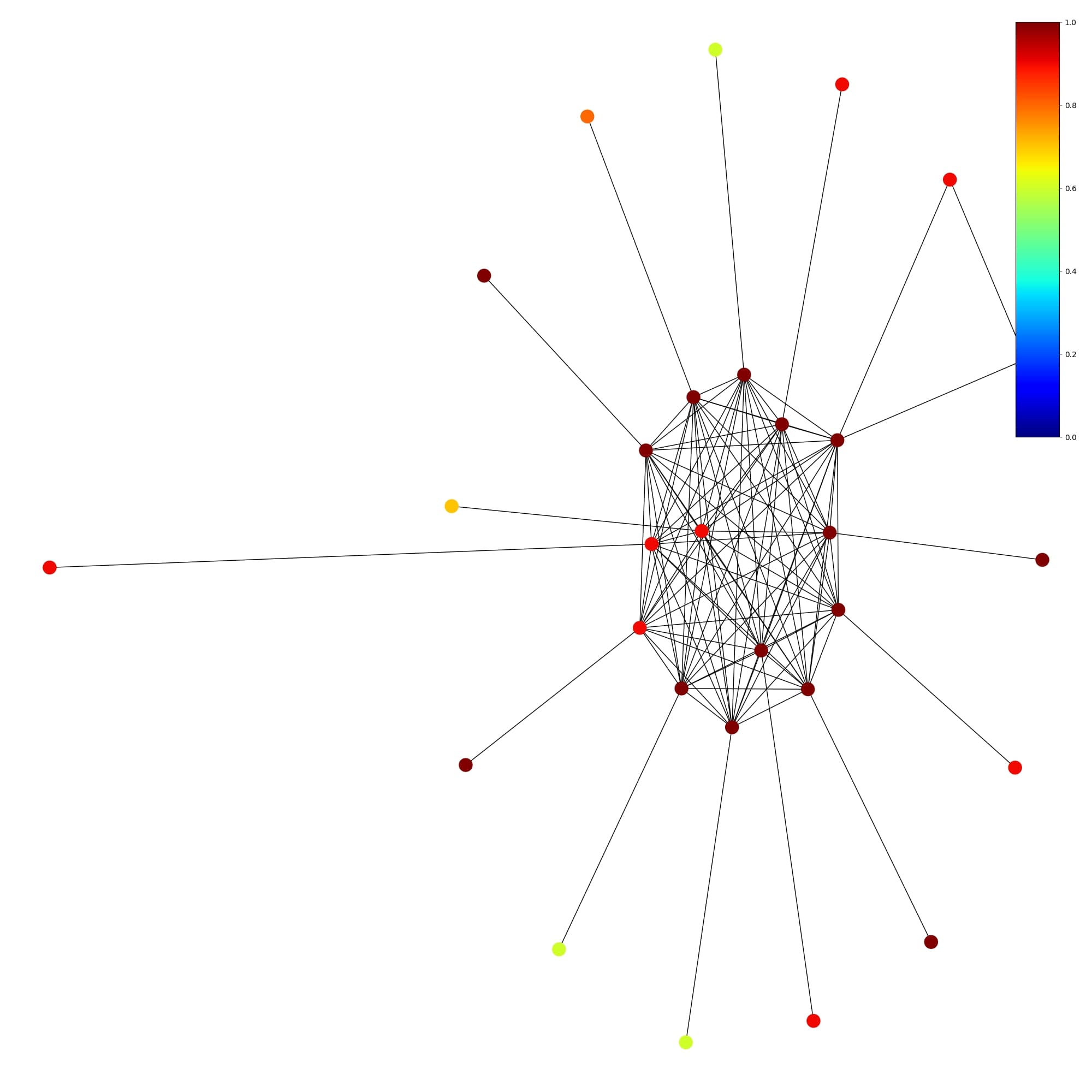}
\caption{Graph Heat Map from Diffusion of ndm.ox.ac.uk.}
\end{figure}
\begin{figure}[H]
\centering
\includegraphics[width=7cm]{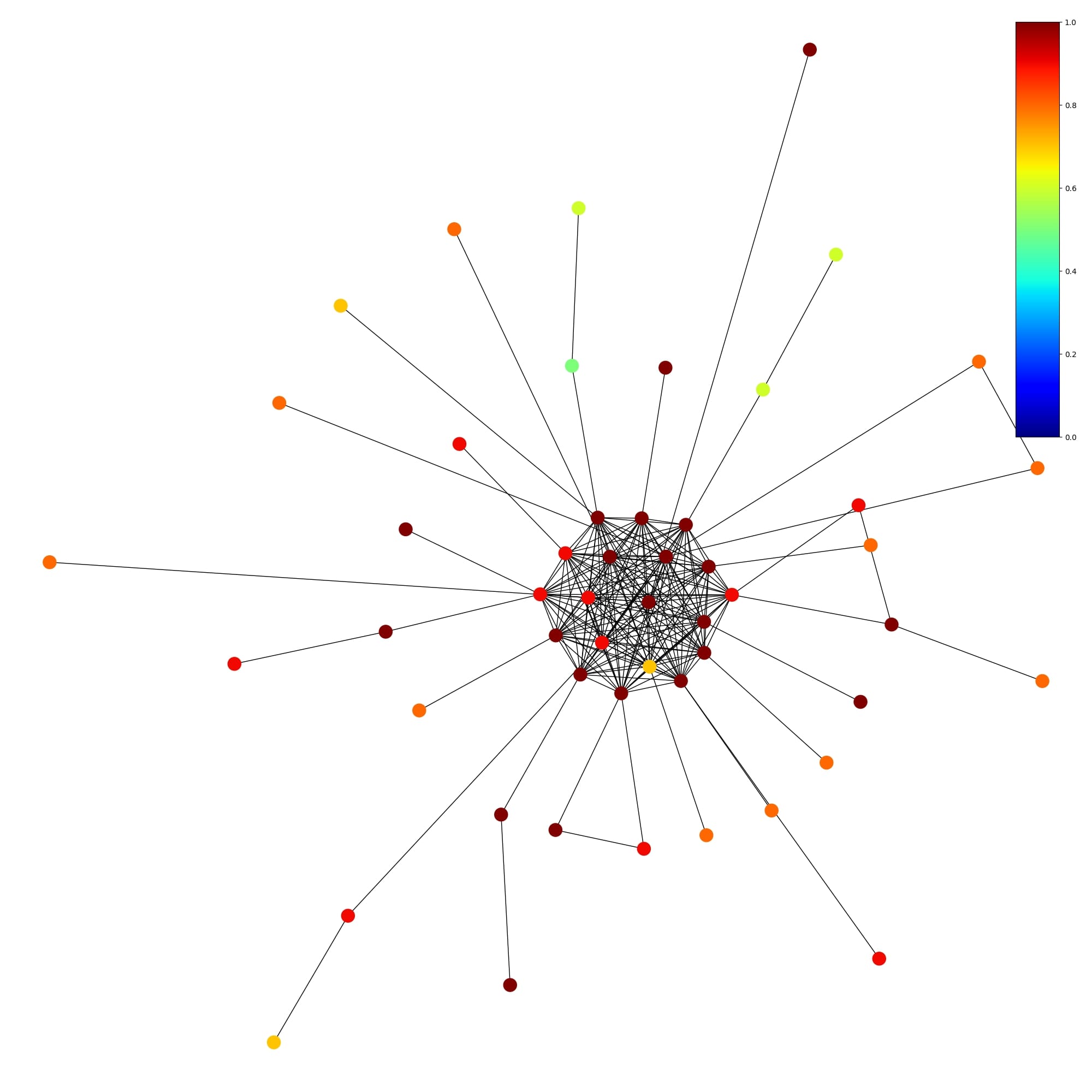}
\caption{Graph Heat Map from Diffusion of cnb.csic.es.}
\end{figure}

\bibliographystyle{IEEEtran}
\bibliography{references}




\end{document}